\documentclass[twocolumn,secnumarabic,amssymb, nobibnotes, aps, prb,groupedaddress,superscriptaddress]{revtex4-2}
\usepackage[utf8]{inputenc}
\usepackage{amsmath,graphicx,latexsym,times,color}
\usepackage{setspace}
\usepackage[hidelinks]{hyperref}
\usepackage{array}
\usepackage{textcomp}
\usepackage{titlesec}
\usepackage{physics}
\usepackage{gensymb}
\usepackage{nccmath}
\usepackage{empheq}
\usepackage{layouts}

\renewcommand{\abs}[1]{\left|#1\right|}
\newcommand{\avg}[1]{\langle#1\rangle}

\DeclareMathOperator{\e}{e}
\DeclareMathOperator{\ssf}{S}
\renewcommand{\i}{\text{i}}
\newcommand{\M}{\mathbf{M}}

\newcommand{\F}{\mathcal{F}}
\newcommand{\U}{\mathcal{U}}
\newcommand{\Ent}{\mathcal{S}}
\newcommand{\Z}{\mathcal{Z}}
\newcommand{\Si}{\mathbf{m}_i}
\newcommand{\Sj}{\mathbf{m}_j}
\newcommand{\Sk}{\mathbf{m}_k}
\newcommand{\Sl}{\mathbf{m}_l}

\renewcommand{\S}{\mathbf{m}}
\newcommand{\kb}{k_\mathrm{B}}
\newcommand{\ncycle}{n_\mathrm{cycle}}
\newcommand{\ei}{\epsilon_I} %
\newcommand{\idxSetZeroMode}{\mathbb{K}_\text{zero}}
\newcommand{\idxSetHarmMode}{\mathbb{K}_\text{harm.}}
\newcommand{\zeromodeSet}{\zeta}

\newcommand{\q}{\mathbf{q}}

\newcommand{\fccMnRe}{{fcc-Mn/Re(0001)}}

\newcommand{\hcpMnRe}{{hcp-Mn/Re(0001)}}

\newcommand{\bzpoint}[1]{$\overline{\text{#1}}$}

\newcommand{\diff}{\mathrm{d}}

\renewcommand{\epsilon}{\varepsilon}
\renewcommand{\phi}{\varphi}

\usepackage[normalem]{ulem}
\usepackage[cmyk,dvipsnames]{xcolor}

\begin{document}
\title{Entropy-driven phase transition in a non-collinear antiferromagnet \\
due to higher-order exchange interactions}
\author{Leo Kollwitz}
\email{kollwitz@physik.uni-kiel.de}
\affiliation{Institute of Theoretical Physics and Astrophysics, University of Kiel, Leibnizstrasse 15, 24098 Kiel, Germany}

\author{Moritz A. Goerzen}
\affiliation{Institute of Theoretical Physics and Astrophysics, University of Kiel, Leibnizstrasse 15, 24098 Kiel, Germany}
\affiliation{CEMES, Universit\'e de Toulouse, CNRS, 29 rue Jeanne Marvig, F-31055 Toulouse, France}

\author{Bjarne Beyer}
\affiliation{Institute of Theoretical Physics and Astrophysics, University of Kiel, Leibnizstrasse 15, 24098 Kiel, Germany}

\author{Hendrik Schrautzer}
\affiliation{Institute of Theoretical Physics and Astrophysics, University of Kiel, Leibnizstrasse 15, 24098 Kiel, Germany}
\affiliation{Science Institute and Faculty of Physical Sciences, University of Iceland, VR-III, 107 Reykjavík, Iceland}

\author{Stefan Heinze}
\affiliation{Institute of Theoretical Physics and Astrophysics, University of Kiel, Leibnizstrasse 15, 24098 Kiel, Germany}
\affiliation{Kiel Nano, Surface, and Interface Science (KiNSIS), University of Kiel, 24118 Kiel, Germany}

\date{\today}

\begin{abstract}
The triple-Q state arises due to the superposition of three symmetry equivalent spin spirals stabilized by higher-order exchange interactions. It has been predicted more than 20 years ago but was only recently discovered in a Mn monolayer on the Re(0001) surface. To date little is known about the thermodynamic properties of this intriguing non-coplanar spin state. Here, we reveal a low-temperature phase transition between the triple-Q and the row-wise antiferromagnetic state in this system via Monte Carlo simulations based on an atomistic spin model parametrized by density functional theory. By modeling the free energy landscape in terms of thermal excitations we derive an analytical expression of the partition function, which allows us to prove that the phase transition is driven by entropy. The predicted phase transition is not unique to Mn/Re(0001) but appears for a wide range of magnetic interaction parameters and is expected to occur also for other multi-Q states.
\end{abstract}
\maketitle

Antiferromagnetic materials have been proposed to play an important role in the future development of spintronic devices~\cite{baltz2018antiferromagnetic}. Compared to established ferromagnetic systems, antiferromagnets come with many benefits, including a large robustness against external magnetic fields, absence of internal stray fields, and ultrafast spin dynamics. Recently,
non-collinear antiferromagnets have moved into
the research focus due to their exceptional
topological and dynamical properties~\cite{Rimmler2024}. A promising material class are chiral antiferromagnets such
as the compounds Mn$_3$Sn and Mn$_3$Pt.
Recently, it has been demonstrated that their
coplanar spin state can be used for room temperature
all-antiferromagnetic
tunnel junctions~\cite{Qin2023,Chen2023}
and that it can be switched by electrical currents
via spin-orbit torque~\cite{Zheng2025}.
Further, non-coplanar spin alignments are
possible in antiferromagnets such as 
the so-called triple-Q (3Q) state which
consists of a superposition of three spin spiral (1Q) states~\cite{kurz2001three}.
In this spin state, the magnetic moments point along the principle axes of a tetrahedron, leading to the presence of topological orbital moments despite a vanishing net magnetization~\cite{hanke2016role,haldar2021distorted} and the absence of spin-orbit coupling. Therefore, the 3Q state exhibits intriguing transport properties in the form of a spontaneous topological Hall effect~\cite{martin2008itinerant,hanke2016role,takagi2023spontaneous,park2023tetrahedral} and can induce topological superconductivity in magnet-superconductor-hybrid
systems \cite{Bedow2020,Nickel2025}.

Recently, the 3Q state was discovered 
in %
a Mn monolayer on the Re(0001) surface using spin-polarized scanning tunneling microscopy 
experiments~\cite{spethmann2020discovery}. After this initial finding two further ultrathin film systems -- %
Pd/Mn and Rh/Mn bilayers on Re(0001)~\cite{nickel2023coupling} -- and also the 
bulk material
Co$_{1/3}$TaS$_2$~\cite{takagi2023spontaneous, park2023tetrahedral} have been experimentally verified to host this intriguing magnetic ground state. 
Even more recently, it has also been observed
in magnetic Mn bilayers on Ir(111) \cite{saxena2024}.
The 3Q state is stabilized in antiferromagnetic films by higher-order exchange
interactions (HOI) such as the biquadratic or the
four-spin interactions
as has been explained based on density functional theory (DFT) calculations and spin models~\cite{kurz2001three, spethmann2020discovery, hoffmann2020systematic,nickel2023coupling,Beyer2025}. It has further been demonstrated that
the 3Q state can be distorted due to magnetocrystalline
anisotropy or topological chiral interactions \cite{nickel2023coupling,haldar2021distorted}. 

While many more complex magnetic ground states, which are stabilized
by HOI,
have been experimentally observed at low temperatures
in ultrathin transition-metal 
films~\cite{heinze2011spontaneous,Yoshida2012,Romming2018,Kroenlein2018, gutzeit2022nano, gutzeit2023spontaneous,Nickel2025b} and have also been predicted for
two-dimensional van der Waals magnets~\cite{Xu2022,Li2023,Pan2024}, not much is known about the thermal properties of these typically non-collinear magnetic states.
It has been previously shown via Monte Carlo (MC) simulations 
that the introduction of HOI into a spin model can cause a non-trivial behavior at low temperatures for a system with a ferrimagnetic ground state~\cite{wieser2008entropy}. However, MC simulations including HOI have so far not been
applied to non-collinear antiferromagnets in 
two-dimensional systems.

Here, we reveal the occurrence of a non-trivial phase transition between the 3Q ground state and the corresponding single-Q 
state, i.e.~the row-wise antiferromagnet, at low temperatures via
MC simulations based on an atomistic spin
model including higher-order interactions. We demonstrate that this low-temperature phase transition is driven by entropy and predict that it can be observed in Mn/Re(0001) at
temperatures of about 30~K
based on using interaction parameters obtained from density functional theory calculations. 
By varying the interaction constants starting from the DFT values we show that the phase transition is a robust result.

The theoretical framework applied in our work to model the film
systems in the relevant temperature regime and
to explain the results from MC simulations
is based on the expansion of the energy landscape around a local energy minimum
in the Hessian eigenbasis of the Hamiltonian, which is successfully used in harmonic transition state 
theory~\cite{bessarab2012harmonic,goerzen2022atomistic}.
Using this approach, we can explicitly calculate thermodynamical state functions such as the free energy and the entropy which are
typically not accessible via MC simulations. This method also
allows to obtain phase boundaries
at a significantly lower computational cost compared to the computationally intense MC simulations.
An extension of this method is developed here to include unstable eigenmodes,
which further widens its applicability and
also improves the agreement between the phase diagrams obtained via the
theoretical model and the MC simulations for the investigated two-dimensional systems.  

\section*{}
\noindent{\large{\textbf{Results}}}\par
\noindent\textbf{Atomistic spin model.}
Understanding the emergence of the 3Q state requires to go beyond the
Heisenberg model of pair-wise exchange. In particular, one needs to 
include higher-order exchange interactions~\cite{hoffmann2020systematic}.
The strengths of these terms are usually small compared to the pair-wise exchange and are therefore often neglected in the theoretical description.
However, they can drastically alter the magnetic ground state of certain material systems by breaking the energy degeneracy of single-Q and multi-Q states. In the past, various experiments 
have already been conducted which reveal the emergence
of such non-trivial magnetic ground states~\cite{heinze2011spontaneous,Yoshida2012,spethmann2020discovery,gutzeit2022nano,nickel2023coupling}. Due to the large number of atoms involved to properly model the experimentally observed behavior of such systems, 
a purely quantum mechanical treatment is not feasible to date. Therefore, we employ
an atomistic spin model, which contains interactions between the magnetic moments derived from a quantum mechanical treatment~\cite{hoffmann2020systematic} with interaction parameters obtained from underlying DFT calculations.
In this work the system is described within the atomistic spin model using the 
classical extended Heisenberg Hamiltonian
\begin{align}
    H =& {-\sum_{i,j}J_{ij}\left(\Si\cdot\Sj\right)} -\sum_i K_i(\mathbf{n}\cdot\Si)^2\notag\\
        &-\sum_{i,j} B_{ij}\left(\Si\cdot\Sj\right)^2 -\sum_{i,j,k}Y_{ijk}\left[\left(\Si\cdot\Sj\right)\left(\Sj\cdot\Sk\right) \right. \notag\\
        &\left.\qquad + \left(\Sj\cdot\Si\right)\left(\Si\cdot\Sk\right) + \left(\Si\cdot\Sk\right)\left(\Sk\cdot\Sj\right)\right] \notag \\
        &-\sum_{i,j,k,l}F_{ijkl}\left[\left(\Si\cdot\Sj\right)\left(\Sk\cdot\Sl\right)\right.\notag\\
        & \left.\qquad + \left(\Si\cdot\Sl\right)\left(\Sj\cdot\Sk\right) - \left(\Si\cdot\Sk\right)\left(\Sj\cdot\Sl\right)\right]\,,\label{eq:full_Heisenberg_Hamiltonian}
\end{align}
where $\S_{i}$ is a unit vector representing the direction of the
magnetic moment at lattice site $i$. The included interactions are the bilinear exchange ($J_{ij}$), the magneto-crystalline anisotropy energy ($K_i$), the biquadratic exchange ($B_{ij}$) as well as the four-spin three-site ($Y_{ijk}$) and the four-spin four-site ($F_{ijkl}$) exchange. The interaction constants 
were taken from previous DFT calculations~\cite{spethmann2020discovery} for a Mn monolayer in fcc stacking on the Re(0001) surface,
denoted as fcc-Mn/Re(0001) in the following
(see "Methods" and Supplementary Table S1 for details). We neglect dipole-dipole interactions as they 
are usually small energy contributions on the order of 0.1~meV/atom 
and can be effectively included into the magnetocrystalline
anisotropy energy (MAE) for film systems%
~\cite{Lobanov2016,Draaisma1988}.
The Dzyaloshinskii-Moriya interaction (DMI) is not taken into account since 
it does not affect the 3Q state as it is a superposition of three collinear row-wise antiferromagnetic (RW-AFM) states. Further, the interaction constants
have been found to be
small compared to the bilinear exchange~\cite{spethmann2020discovery, nickel2023coupling},
so that the DMI affects distortions of the ideal 3Q state on a minor scale.

\noindent\textbf{Monte Carlo simulations.}
\begin{figure}
    \centering
    \includegraphics[width=\columnwidth]{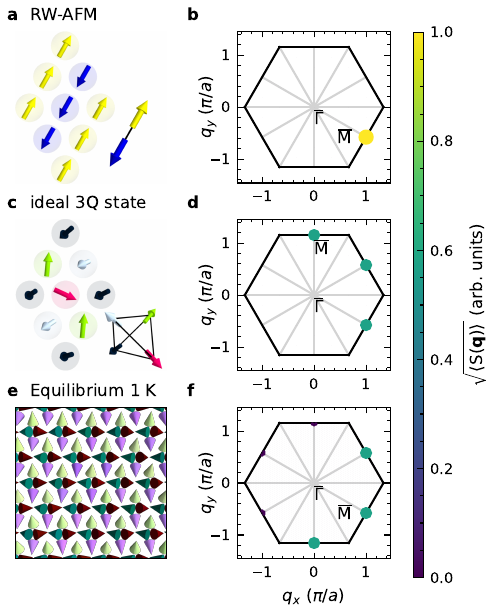}
    \caption{\textbf{Representations of RW-AFM and 3Q state in real and reciprocal space.}
    Schematic real space representation of \textbf{a} the RW-AFM on a hexagonal lattice. This single-Q state shows only one prominent component in its spin-structure factor at one of the \bzpoint{M}-points in the first Brillouin zone \textbf{b}. For the non-coplanar ideal triple-Q state \textbf{c} the spin structure factor shows equally prominent components at all three \bzpoint{M}-points \textbf{d}. In \textbf{e}, \textbf{f} the equilibrium state at $T=1$\,K obtained via MC simulations as described in the main text and the averaged spin structure factor at this temperature are shown.}
    \label{fig:3q_RWAFM_SSF_Brillouin}
\end{figure}
In the case of a Mn monolayer in fcc stacking on the Re(0001)
surface
the 
DFT calculations
reveal a %
small energy difference of only $E_\text{3Q} - E_\text{RW-AFM} = -0.7$\,meV/Mn atom
between the triple-Q and the RW-AFM state~\cite{spethmann2020discovery}.
Due to this flat energy landscape, 
we first verify that the magnetic ground states are consistent between the atomistic spin model and the DFT total energy calculations. The ground state within the atomistic spin model is determined using classical Monte Carlo simulations (see "Methods" for details).
Distinguishing the presence of a triple-Q state from the RW-AFM in the obtained data is not trivial, as both show a vanishing net magnetization.
While quantities such as the scalar-spin chirality defined by
$\chi_{ijk}=\S_i \cdot \left(\S_j\times\S_k\right)$ can be used to distinguish non-coplanar from coplanar states, it fails to also identify the disordered paramagnetic phase. In this work the so-called spin structure factor (SSF) of a magnetic configuration $\M$, defined as
 \begin{align}
    \ssf[\M](\q) = {1\over N^2}\sum_{\alpha\in\{x,y,z\}}\abs{\sum_{i=1}^{N} m_i^\alpha \e^{-\i\q\cdot\mathbf{r}_i}}^2\,,\label{eq:ssf_definition}
 \end{align}
 is used as a general antiferromagnetic order parameter~\cite{gvozdikova2005monte}.
 Here, $N$ is the number of atoms present in the simulation box. This quantity shows the contributions of the spin spiral states
 with wave vector $\mathbf{q}$
 to the magnetic state 
 and corresponds to the intensity measured in inelastic neutron scattering experiments~\cite{rozsa2016complex}.
 Its components at the reciprocal lattice vectors $\mathbf{q}_i,\ (i=1,2,3)$,
 located at the three $\overline{\mathrm{M}}$-points of the first Brillouin zone (BZ),
 measure the presence of each of the three symmetry-equivalent RW-AFM states in a given state. This makes it a suitable order parameter for the distinctions between the 
 RW-AFM (single-Q) state, showing only one prominent
 peak
 on one of the \bzpoint{M}-points, the triple-Q state, which %
 shows peaks at all three distinct \bzpoint{M}-points with reduced intensity (Fig.~\ref{fig:3q_RWAFM_SSF_Brillouin}a-d), and the paramagnetic phase with nearly vanishing components~\cite{gvozdikova2005monte}.
\begin{figure*}
    \centering
    \includegraphics{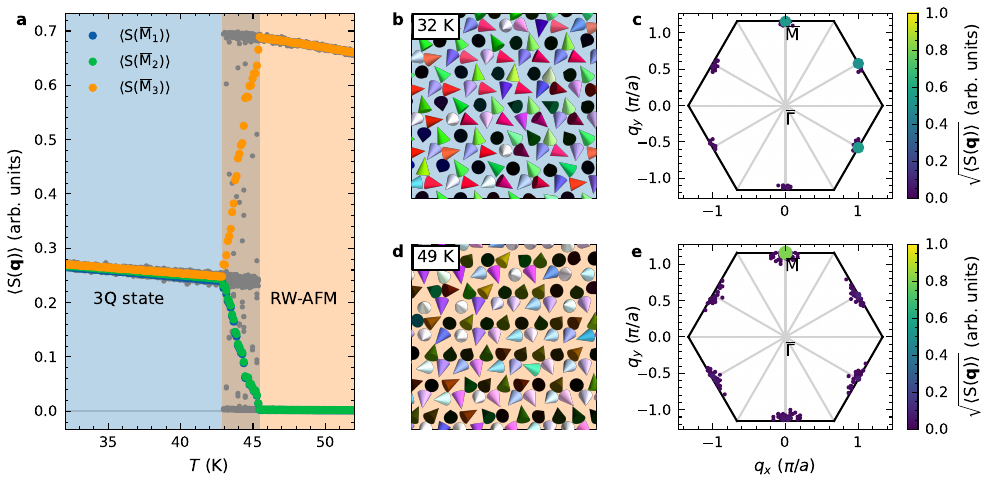}
    \caption{\textbf{Phase transition from the 3Q to the RW-AFM state.}
    \textbf{a} Temperature dependence of the SSF contributions at the three different \bzpoint{M}-points in the first BZ. The colored data points are averaged across 20 independent simulations neglecting MAE. For the lower temperatures (blue background) the three SSF contributions are of nearly equal magnitude, while for higher temperatures (orange background) there is only a single prominent component visible. \textbf{b}, \textbf{d}
    Exemplary equilibrium states at $T=32$\,K and $T=49$\,K showing a 3Q and RW-AFM structure respectively. 
    \textbf{c}, \textbf{e}
    Averaged components of the SSF across the entire first BZ. at $T=32$\,K and $T=49$\,K respectively.}
    \label{fig:transition_SSF}
\end{figure*}
With the DFT interaction parameters from Ref.~\cite{spethmann2020discovery},
for now neglecting the MAE, the equilibrium state at $T=1$\,K 
(Fig.~\ref{fig:3q_RWAFM_SSF_Brillouin}e) indeed shows tetrahedron angles of neighboring spins, characteristic for the ideal triple-Q state. The average components of the spin structure factor (Fig.~\ref{fig:3q_RWAFM_SSF_Brillouin}f) also show prominent peaks on all three \bzpoint{M}-points, with small thermal noise visible. The state was obtained using a simulated annealing approach. For a $60\times 60$ simulation box the simulations start from a uniform random spin configuration at $T=1500\,$K. Using $10^8$ single spin updates for the equilibration process at each temperature, the system is annealed to $T=1$ K using a total of 150 linearly spaced temperature %
steps. The obtained state at $T=1$ K is assumed to be in the vicinity of the ground state and is then minimized %
with respect to its energy using the L-BFGS algorithm \cite{nocedal1999numerical, ivanov2021fast} to remove thermal fluctuations, which indeed results in the ideal triple-Q alignment.

For the simulation at finite temperatures, the temperature dependence of the SSF %
at the three distinct \bzpoint{M}-points in the first Brillouin zone is shown in Fig.~\ref{fig:transition_SSF}a. The data was obtained from 20 independent MC simulations starting at $T=52$ K from a previously equilibrated state. The system was annealed to $T=32$ K using a total of 200 linearly spaced temperature steps. For each temperature, $4.32\cdot10^7$ single spin updates were performed for the equilibration. After that, $2.5\cdot10^4$ samples across $1.8\cdot10^8$ single spin updates were taken for the computation of the averages. The averaged data of each individual simulation
was then averaged again across the 20 independent simulations. For low temperatures $T<43$\,K the \bzpoint{M}-point components of the SSF are of roughly equal magnitude,
indicating the presence of
the triple-Q state. Due to the averaging, this could have also been caused by the presence of three RW-AFM domains or the switching from one RW-AFM orientation to another during the simulation. However,
the real space representation of exemplary states at different temperatures shows the presence of the triple-Q state 
(Fig.~\ref{fig:transition_SSF}b,c) in accordance with the determined ground state. Across the temperature range $43\,\mathrm{K}\leq T \leq 46\,\mathrm{K}$ the system undergoes a phase transition between a triple-Q phase present at lower temperatures and the corresponding single-Q (RW-AFM) state present at higher temperatures 
(Fig.~\ref{fig:transition_SSF}d,e). This transition shows the complex and non-trivial effects that %
HOI can have %
on the free energy landscape of the magnetic configurations in %
Mn/Re(0001).

\begin{figure}
    \centering
    \includegraphics[width=\columnwidth]{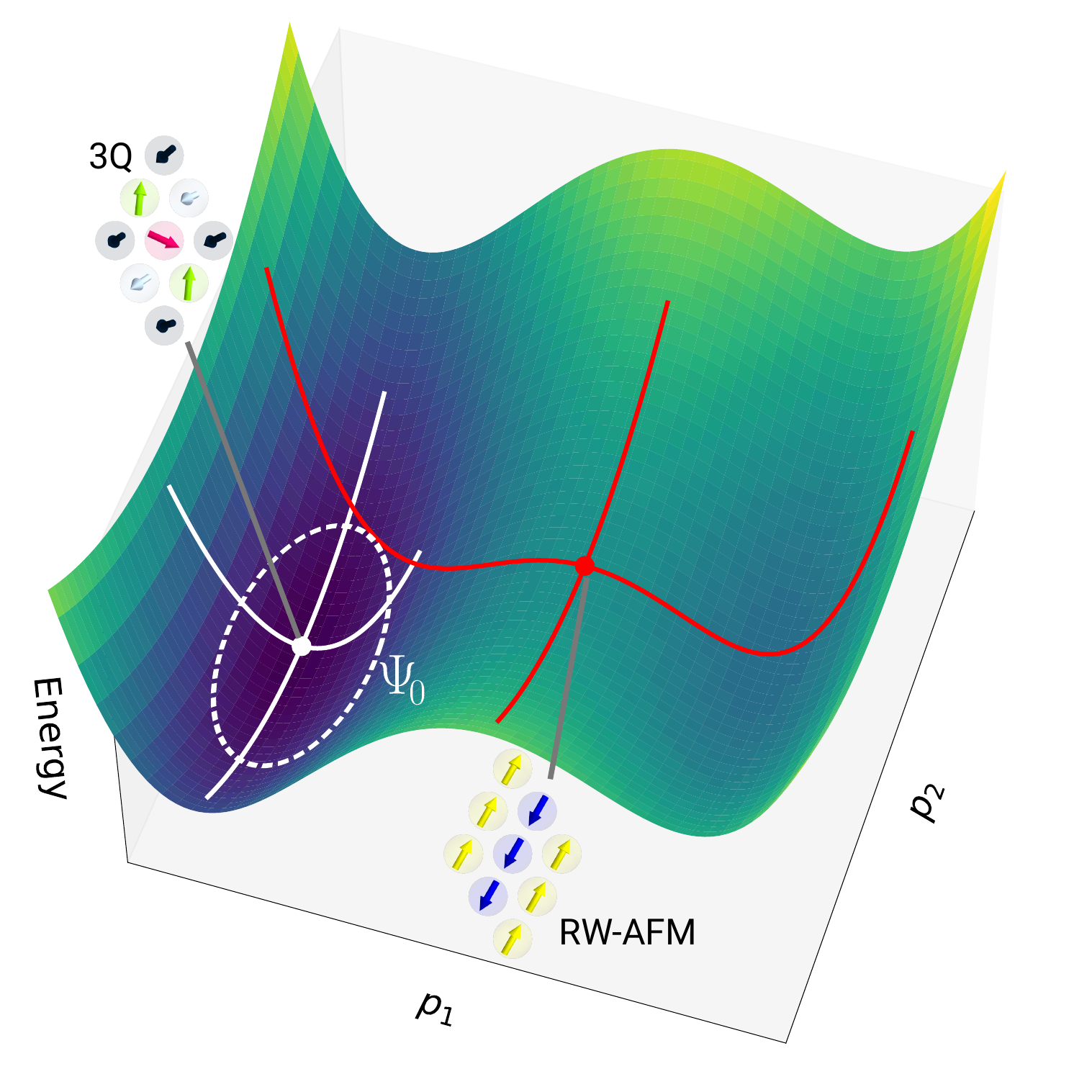}
    \caption{\textbf{Energy landscape around the 3Q and the RW-AFM state.}
    Illustration of the energy landscape using different kinds of Hessian eigenmodes around a local energy minimum (white) or a general stationary point with a negative
    eigenvalue (red). In the case of \fccMnRe\ these correspond to the
    3Q and the RW-AFM state.}
    \label{fig:mode_fit}
\end{figure}

\noindent\textbf{Free energy computation.}
In order to theoretically describe the behavior at finite temperatures of 
Mn/Re(0001)
and to quantify it in terms of the relevant thermodynamic state functions, which here are the (Helmholtz) free energy $\F$, the entropy $\mathcal{S}$ and the internal energy $\mathcal{U}$, an analytical expression for the partition function $\Z$ %
of an ensemble of magnetic configurations is needed. This can be achieved within the framework of the harmonic approximation (HA) of the energy landscape around some local energy minimum state. Here, we will give a brief outline of this framework, which is described in detail in Ref.~\cite{goerzen2024PhD}.
In a second order approximation to the Hamiltonian 
(Eq.~\eqref{eq:full_Heisenberg_Hamiltonian}) 
magnetic excitations can be expressed in the eigenbasis of its Hessian, which contains the second derivatives of the energy with respect to variations of the magnetic configuration. The corresponding eigenvectors are commonly divided into harmonic modes and zero modes, which are characterized by their respective eigenvalue $\lambda>0$ or $\lambda=0$.
The harmonic
modes are modeled by a quadratic energy dispersion along their normal coordinate $p$, while the zero modes -- corresponding to symmetries of the Hamiltonian -- explicitly contribute with their entire %
configuration space volume $\zeta$. This leads to the approximate expression of the partition function for
an ensemble of configurations $\M$ in the subspace $\Psi_0$ around a local energy minimum with energy $E_0$, which reads
\begin{align}
    \Z^\text{HA} = &\int_{\Psi_0}\e^{-\beta E(\M)}~\diff{\M}\\
    \approx & \left(\prod_{n\in \idxSetZeroMode}\zeromodeSet_n\right) \left(\prod_{n\in \idxSetHarmMode}\int_{-\infty}^{\infty}\e^{-\frac{1}{2}\beta\lambda_n p_n^2}\,\diff p_n\right)\e^{-\beta E_0} \label{eq:partition_function_ha_2}\\
     = & \left(\prod_{n\in \idxSetZeroMode}\zeromodeSet_n\right) \left(\prod_{n\in \idxSetHarmMode}\sqrt{\frac{2\pi}{\beta\lambda_n}}\right)e^{-\beta E_0}\\
     = & \Z_\mathrm{zero}\cdot \Z_\mathrm{harm.}\cdot e^{-\beta E_0}%
\end{align}
with the usual $\beta=1/\kb T$. 
Here $\idxSetZeroMode$ and $\idxSetHarmMode$ are the index sets of the zero and harmonic modes respectively. Using the
harmonic approximation, 
we are therefore able to give quantitative entropy differences between the involved RW-AFM and triple-Q states. Surprisingly, computing the eigenvalues of the Hessian of the RW-AFM state reveals the presence of unstable eigenmodes, %
characterized by negative eigenvalues $\lambda_n<0$. Thus, the state is not a local 
energy minimum, even though it appears as the thermal equilibrium state at finite temperatures. %

We can further extend %
the partition function
to describe
general stationary points with some number of negative eigenvalues in its Hessian instead of only local energy minima (cf. Fig.~\ref{fig:mode_fit}). Using numerical mode following algorithms~\cite{muller2018duplication, malottki2025eigenmode} we can identify two different kinds of unstable modes present in the system. The first kind belonging to the lowest eigenvalues of the Hessian are unstable periodic modes, which are modeled by the dispersion
\begin{align}
    E(p)  = -2\ei \sin^2\left(\frac{1}{2}p\sqrt{\frac{\abs{\lambda}}{\ei}}\right)\, \label{eq:neg_periodic_dispersion}
\end{align}
with the corresponding eigenvalue $\lambda<0$ and the absolute inflection point energy $\ei>0$ of the dispersion. This energy is defined with respect to $E(p=0)=0$.
It is determined using mode following algorithms as the energy of the state along the normal coordinate, at which the corresponding eigenvalue $\lambda$ vanishes.
The second kind of unstable mode is modeled using the quartic dispersion
\begin{align}
    E(p) = \frac{1}{2}\lambda p^2 + \frac{5\lambda^2}{144\ei}p^4\,. \label{eq:neg_quartic_dispersion}
\end{align}
Again, the different modes are assumed to be independent of one another, so that we can write the partition function in the extended harmonic approximation (EHA) as a product of integrals over the Boltzmann distribution of the corresponding energy dispersions as
\begin{align}
    \Z^\text{EHA} &= \prod_\text{modes}\left(\int_{\text{period}} \e^{-\beta E_\text{mode}(p)}\diff p\right)\e^{-\beta E_0}\\
    &= \Z_\mathrm{zero}\cdot \Z_\mathrm{harm.} \cdot \Z_\mathrm{per.}\cdot \Z_\mathrm{quart.}\cdot \e^{-\beta E_0}\,.\label{eq:total_Z_harm_approx_product_contr}
\end{align}
Zero and unstable periodic modes have a finite period, over which the integration is performed, while the harmonic and quartic modes are assumed to
be non-periodic, thus the integration is performed over the interval $(-\infty,\,\infty)$. 
For the modeling of the individual modes and their contributions to the partition function see "Methods". This expression for the partition function lets us derive the %
free energy $\F=-\ln\left(\Z\right)/\beta$, the internal energy $\U=-\partial_\beta\ln{\Z}$ and the entropy $\Ent=\left(\U-\F\right)/T$ (for explicit formulas see Supplementary Note~S1)
of different initial configurations such as the RW-AFM and 
the triple-Q state.

\begin{figure}
    \centering
    \includegraphics{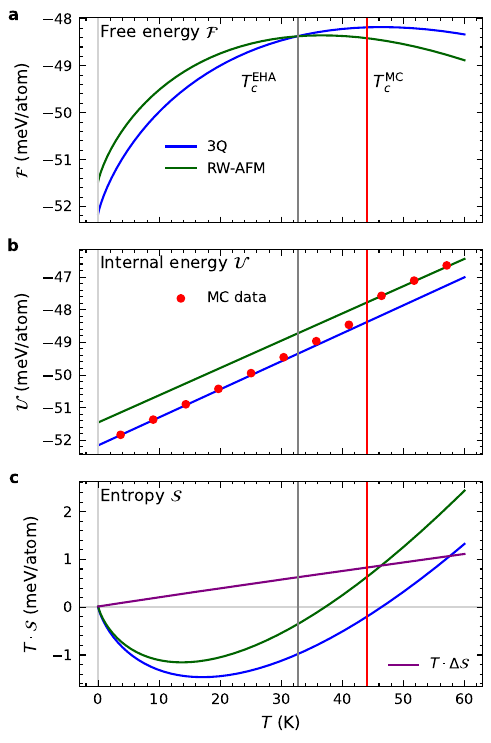}
    \caption{\textbf{Thermodynamic state functions across the phase transition.}
    Temperature dependence of \textbf{a} the free energy $\mathcal{F}$, \textbf{b} internal energy $\mathcal{U}$ and \textbf{c} (temperature scaled) entropy $\mathcal{S}$ for the triple-Q state and RW-AFM in fcc-Mn/Re(0001) computed within the extended harmonic approximation. The blue (green) curves show the thermodynamic state functions for the triple-Q (RW-AFM state). The vertical lines indicate the transition temperatures obtained within the extended harmonic approximation (gray, $T_c^\text{EHA}=32.7$ K) and the MC simulations (red, $T_c^\text{MC}=44.0$ K).
    Additionally, data of the internal energy $\U$ obtained from the MC simulations is shown for exemplary temperatures in the triple-Q as well as RW-AFM phase. In \textbf{c} the (temperature scaled) entropy difference $T\cdot\left(\Ent_\text{RW-AFM}-\Ent_\text{3Q}\right)$ is shown in purple.}
    \label{fig:harmonic_approx_results}
\end{figure}

 Treating the different kinds of modes in the EHA 
 allows us to quantitatively compute the thermodynamic state functions of the triple-Q 
 and the RW-AFM state 
 (Fig.~\ref{fig:harmonic_approx_results}). 
 At low temperatures, the triple-Q state exhibits a lower free energy than the RW-AFM state (Fig.~\ref{fig:harmonic_approx_results}a), thus making it the thermodynamically preferred state.
 However, at $T_c^{\text{EHA}}=32.7$\,K the two curves cross and the RW-AFM becomes favorable compared to the triple-Q state, albeit its larger 
 energy (Fig.~\ref{fig:harmonic_approx_results}b). Due to the small energy difference between the two states, this transition happens at a temperature well below the Néel temperature, which has been determined to be $T_N\approx115$\,K (see Supplementary Note~S2 and Supplementary Figure~S1). 
 Comparing the predicted internal energy $\U$ with the one sampled during the MC simulations 
 (Fig.~\ref{fig:harmonic_approx_results}b) shows a very good agreement for the triple-Q as well as the RW-AFM phase with the
 extended harmonic approximation. 
 For low temperatures, the internal energy in both cases has a nearly linear dependence on the temperature as expected from the equipartition theorem.
 Computing the entropy of the two states 
 (Fig.~\ref{fig:harmonic_approx_results}c) shows the RW-AFM being entropically favored. 
 
 This finding is consistent with a previous theoretical study \cite{henley1989ordering} showing single-Q states being favored compared to multi-Q states in the presence of thermal fluctuations. In our case, however, due to the presence of HOI, there is an energetic difference between the triple-Q state and RW-AFM, making the former the equilibrium state at sufficiently low temperatures. 
 The entropy difference at $T_c^{\mathrm{EHA}}=32.7$\,K between the two states is calculated to be $\Delta\Ent=1.92\cdot10^{-2}\,\mathrm{meV}\cdot\mathrm{K}^{-1}/\mathrm{atom}$. 
 Note, that for both states the entropies diverge for $T\rightarrow 0$, apparently violating the third law of thermodynamics, which is due to
 the harmonic approximation not being defined at absolute zero. However, the entropic energy contribution $T\Ent$ has a finite limit, which is zero if only harmonic modes are present. 

\begin{figure}
    \centering
    \includegraphics{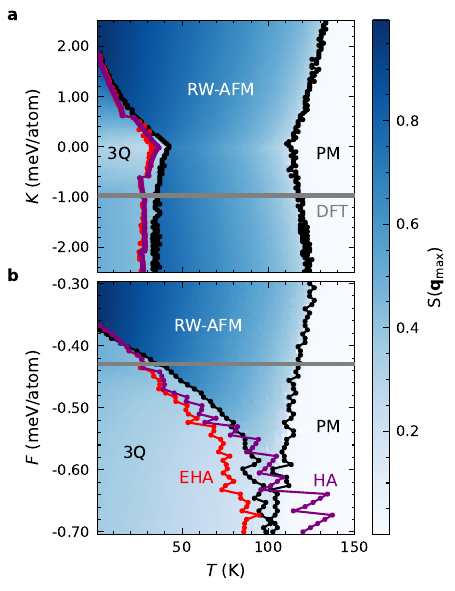}
    \caption{\textbf{Phase diagrams vs.~magnetic interaction strengths.}
    Phase diagrams showing the maximum component of the SSF (see color bar) for different interaction parameters of \textbf{a} the MAE and \textbf{b} the four-spin four-site exchange. The horizontal gray lines indicate the value of the interaction parameters obtained via DFT in Ref.~\cite{spethmann2020discovery} for fcc-Mn/Re(0001). The black lines show the phase boundaries obtained in the Monte Carlo simulations. The purple (red) phase boundaries between the RW-AFM and triple-Q state are obtained within the (extended) harmonic approximation excluding (including) unstable degrees of freedom.
    }
    \label{fig:phase_diagrams}
\end{figure}

So far, we have focused on the magnetic interaction parameters
obtained via DFT for a Mn monolayer in fcc stacking on Re(0001) neglecting the magnetocrystalline anisotropy.
However, the observed phase transition is 
stable across a wide range of interaction parameters
and allowing critical temperatures of up to 100~K
as we have found by varying the interaction strengths 
in our MC simulations (Fig.~\ref{fig:phase_diagrams}).
In particular, the temperature dependence of the largest component of the SSF is shown for varying interaction constants of the MAE 
(Fig.~\ref{fig:phase_diagrams}a)
and the four-spin four-site exchange 
(Fig.~\ref{fig:phase_diagrams}b). For the RW-AFM (single-Q)
state this component will be close to unity, while for the ideal triple-Q state it is about $1/3$. In the disordered paramagnetic (PM) phase the component will be close to zero, allowing for a clear visual difference between the three relevant phases. The obtained phase boundaries from the MC simulations are shown in black.

We find that the phase transition from the 3Q to the RW-AFM state at low
temperatures, which we found for vanishing MAE (Fig.~\ref{fig:transition_SSF}), 
is robust upon changing the value of the MAE (Fig.~\ref{fig:phase_diagrams}a).
There is also a very good agreement between the extended harmonic approximation (red), in which the unstable modes in the RW-AFM phase are considered as described above, and the MC simulations for both easy-plane ($K<0$) and easy-axis ($K>0$) anisotropies. Notable here is the varying magnitude of $\ssf$ in the triple-Q phase for different values of $K$. 
We find that the presence of the MAE in this particular system leads to a large distortion of the ground state towards a 2Q (1Q) state for easy-plane (easy-axis) anisotropies (see Supplementary Note~S3 and Supplementary Figures S2-S4). A similar effect, present here within the atomistic model of \fccMnRe, has been predicted by DFT calculations to also occur in the \hcpMnRe\ system~\cite{haldar2021distorted}. However, this distortion does not qualitatively affect the transition to the RW-AFM phase. When ignoring the contribution from the unstable modes
for the RW-AFM state, the purple phase boundary is obtained. In the case of varying MAE, the difference is effectively negligible. This is a result of the small number of unstable modes, which is nearly independent of the value of $K$.

For the four-spin four-site exchange 
(Fig.~\ref{fig:phase_diagrams}b) the extended harmonic approximation captures not only the qualitative shape well, but also the quantitative magnitude of the transition temperatures across various parameter strengths and temperature ranges. In contrast to the MAE, lower values of the four-spin four-site interaction strength $F$ shift the transition towards higher temperatures. This can be explained by an increased stability of the triple-Q state compared to the RW-AFM state, which manifests itself in the increasing number of unstable modes present in the latter 
(see Supplementary Note~S4 and Supplementary Figure~S5). This is also visible in the phase boundaries obtained within the HA. Neglecting the contribution of the unstable modes (purple) for low values of $F$ leads to an overestimation of the transition temperature, which might even surpass the Néel temperature marking the transition to the disordered paramagnetic phase. 

A jagged shape of the purple phase boundary is visible at higher
strength of the four-spin four site interaction (Fig.~\ref{fig:phase_diagrams}b)
which is caused by discontinuous jumps of the computed transition temperature. These are a result of possibly degenerate eigenvalues of the Hessian of the RW-AFM state
abruptly changing their sign with decreasing $F$. This results in an abrupt change in the number of considered harmonic modes for the purple phase boundary. Comparing this behavior with the extended harmonic approximation (red), which includes the contribution of unstable modes, a much better agreement with the Monte Carlo simulations can be seen.
Although there are still notable fluctuations in the obtained transition temperature, they are much less pronounced. These fluctuations as well as the general discrepancy to the Monte Carlo simulations are a result of the
approximation of the energy landscape, which may
deviate significantly from the original
at larger thermal energies. Due to the resulting distortions the assumption of independent modes might also be no longer justified. Further, a changing sign of some highly degenerate eigenvalues still leads to a drastic change in the modeling of the respective modes. Nonetheless, the harmonic approximation with the inclusion of unstable modes captures the trend of the phase boundaries between the triple-Q and the RW-AFM state
very well, especially in the low temperature regime $T\lesssim50$ K.

\section*{}
\noindent{\large{\textbf{Discussion}}}\par

Based on atomistic Monte Carlo simulations for the Mn monolayer on a Re(0001) substrate using a spin model parameterized from DFT, we predict the occurrence of a low temperature phase transition between the triple-Q ground state and the RW-AFM state. This phase transition, which occurs in
addition to that to the paramagnetic phase at the N\'eel 
temperature,
is induced by the presence of HOI, which leads to a splitting of the energy degeneracies in the bilinear exchange interactions between the RW-AFM and the triple-Q state. Our results show the crucial influence of the HOI in the finite temperature behavior of 
Mn/Re(0001) in addition to causing a non-coplanar multi-Q ground state. This highlights the importance of the consideration of HOI despite their small energy scale compared to the bilinear exchange.

Modeling the energy landscape in the vicinity of the different spin states, we are able to quantitatively compute all relevant thermodynamic state functions, which are in general not accessible using standard Monte Carlo techniques.  The resulting phase boundaries are in very good agreement with Monte Carlo simulations not only for the initial parameter set obtained from DFT, but across a wide range of different interaction strengths as well, while requiring only a small fraction of the computational cost. This for one showcases the robustness of the predicted phase transition against the initial parametrization, but also highlights the accuracy of our theoretical model. 
By going beyond the harmonic approximation, we are able to include states containing unstable Hessian eigenmodes, which still play an important role in the thermodynamic behavior of this system, as seen from the occurrence of the RW-AFM state as the equilibrium state at finite temperatures. The extended harmonic approximation provides useful insights to the finite temperature behavior of a system by allowing for a fast comparison of the thermodynamic state functions of different relevant states, which are usually already known from prior DFT studies. 
This model can also be easily extended to include other kinds of interactions in the Hamiltonian e.g.~topological-chiral
interactions \cite{Grytsiuk2020,haldar2021distorted}.

Based on our investigation, this first order phase transition between the triple-Q ground state and the entropically favored RW-AFM state can occur, if these two states are energetically close to each other, such that $\Delta \U / \Delta \Ent < T_c$, with the critical temperature $T_c$ marking the transition to the disordered paramagnetic phase. 
The triple-Q state has also been experimentally observed
in a Pd/Mn bilayer on Re(0001)~\cite{nickel2023coupling}.
However, in this system the condition is not fulfilled since
the energetic difference to the RW-AFM state obtained via
DFT~\cite{nickel2023coupling} is by far too large for an additional phase transition to occur, as
we have confirmed by MC simulations as well as the harmonic approximation (see Supplementary Note~S5 and Supplementary Figures~S6 and S7).

Previously, a first-order phase transition has been reported from a ferrimagnetic to a N{\'e}el state that
can be induced in the presence of HOI~\cite{wieser2008entropy}
using DFT parameters obtained for
a free-standing (unsupported) V monolayer~\cite{Kurz_Phd_thesis}.
It was argued that an entropic difference between the 
two states
is the origin of the transition~\cite{wieser2008entropy}, however, 
without providing calculations of the free energy or the entropy.
Our work, on the other hand, is based on realistic magnetic interaction strengths from an underlying DFT parametrization of an already experimentally realized ultrathin film system with a non-collinear magnetic ground state~\cite{spethmann2020discovery,spethmann2021discovery}.
Further, the entropy-driven mechanism of the transition
is explicitly demonstrated by the extended harmonic approximation.
Note, that a similar phase transition as predicted in our work
for two-dimensional systems
has recently been observed experimentally in the bulk van-der-Waals material Co$_{1/3}$TaS$_2$ using transport and neutron scattering measurements~\cite{takagi2023spontaneous, park2023tetrahedral}.

For the \fccMnRe\ film system, there is a discrepancy in the magnetic ground state between the DFT total energy calculations for the pseudomorphic Mn monolayer
and SP-STM experiments, which show a RW-AFM ground state \cite{spethmann2020discovery}. Recently, it has
been demonstrated that this difference is due to the fact that
the Mn monolayer actually exhibits a lateral shift with 
respect to the surface \cite{zahner2024kicking} that 
had not been considered before. 
Upon taking this structural effect into account in DFT calculations the RW-AFM state is energetically lower than 
the 3Q state
in agreement with experiments \cite{zahner2024kicking}.

On the other hand, the closely related \hcpMnRe\ film system exhibits an experimentally verified 3Q ground state~\cite{spethmann2020discovery}. While the DFT total
energy calculations are consistent with this experimentally
observed ground state, the
DFT parametrization of the atomistic spin model (Eq.~(1))
suffers from large frustration effects and a shallow
energy landscape in the explicit consideration of HOI. Therefore, in MC simulations we do not obtain 
the 3Q state at the lowest temperature.
Nevertheless, since the phase transition from 
the 3Q to the RW-AFM state 
is robust against changes in the parametrization of the spin model and \hcpMnRe\ possesses an even smaller total energy difference 
$\Delta \U$
between the 3Q and the RW-AFM state in the DFT 
calculations~\cite{spethmann2020discovery}
than \fccMnRe\, it is a promising candidate to experimentally verify the
intriguing non-trivial behavior at low temperatures
predicted here for ultrathin antiferromagnetic films
with a 3Q ground state, e.g.~using SP-STM at variable 
temperatures~\cite{Sessi2009}. Based on our work, 
we further expect that a similar
phase transition can occur in two-dimensional magnets
from other multi-Q states, e.g.~the $\uparrow\uparrow\downarrow\downarrow$ states \cite{Romming2018,Kroenlein2018},
to their corresponding single-Q states.

\section*{}
\noindent{\large{\textbf{Methods}}}\par

\noindent\textbf{Magnetic interaction constants of the spin model.}
For the interaction constants of the atomistic spin model given
by Eq.~(1), we have chosen the parameters obtained via DFT
calculations for a Mn monolayer in fcc stacking on the Re(0001)
surface (denoted as fcc-Mn/Re(0001)) given in Ref.~\cite{spethmann2020discovery} and summarized in 
Supplementary Table~S1. In these DFT
calculations the 3Q state is found as the energetically lowest state. 
Experimentally, the corresponding 1Q state, i.e.~the RW-AFM state, has been
observed using spin-polarized scanning tunneling microscopy
(SP-STM)~\cite{spethmann2020discovery}. This discrepancy between theory
and experiment has only recently been resolved \cite{zahner2024kicking}.
It has been demonstrated that there is a structural lateral
shift of the Mn monolayer
in the RW-AFM state which lowers its energy below that of the 3Q
state leading to a consistent picture of theory and experiment.

For a Mn monolayer in hcp stacking on the Re(0001) surface 
(denoted as hcp-Mn/Re(0001)) a
3Q state has been observed in the SP-STM experiments of
Spethmann {\it et al.}~\cite{spethmann2020discovery}. DFT
calculations for hcp-Mn/Re(0001) also obtain the 3Q state as
energetically lowest with respect to all 1Q (spin spiral) 
states~\cite{spethmann2020discovery}.
However, based on the atomistic spin model
parameterized from these DFT calculations we could not stabilize
the 3Q state at low temperatures in Monte Carlo simulations. We
attribute this result to the strong frustration of higher-order
exchange interactions in this system which can be seen from the
small energy difference between the 3Q and 1Q state on the order
of 1~meV/Mn atom found in DFT~\cite{spethmann2020discovery}.

\noindent\textbf{Monte Carlo simulations.}
In order to simulate the system at finite temperatures, we employ classical Monte Carlo simulations using a single-spin updating scheme. For the spin flip mechanism, an adaptive algorithm proposed in Ref.~\cite{alzate2019optimal} has been employed, ensuring an acceptance probability of approximately $50\%$ for each temperature after equilibration. The acceptance probability $A$ to transition from a state $\M$ to the state $\M'$ for each Monte Carlo step is that of the Metropolis algorithm
\begin{align}
    A(\M\rightarrow \M') = \min\left\{1, \e^{-\beta(E(\M')-E(\M))}\right\}\,. \label{eq:acceptance_metropolis}
\end{align}

The Monte Carlo simulations consist of an initial equilibration of the system at each temperature step. After thermal equilibrium is reached, the measured quantities are computed
after every $\ncycle>N$ single spin updates to reduce correlation between them, with $N$ being the number of spins in the simulation box. The mean of the obtained quantities is then taken as the thermal average $\avg{\dots}$. For increased accuracy, a number of statistically independent simulations differing in the seed value of the used Mersenne-twister random-number-generator
are performed. The thermal average is then taken as the arithmetic mean of the averages obtained from each individual simulation.

\noindent\textbf{Harmonic approximation.}\label{subsec:harmonic_approx}
The individual contributions of the periodic unstable modes can be computed as
\begin{align}
    & w\int_{\mathrm{1~period}}\e^{2\beta\ei \sin^2\left(\frac{1}{2}p\sqrt{\frac{\abs{\lambda}}{\ei}}\right)}\,\diff p \\
    &= w\int_{0}^{2\pi\sqrt{\frac{\ei}{\abs{\lambda}}}}\e^{2\beta\ei \sin^2\left(\frac{1}{2}p\sqrt{\frac{\abs{\lambda}}{\ei}}\right)}\,\diff p\\
    &=2w\sqrt{\frac{\ei}{\abs{\lambda}}}\int_{0}^{\pi}\e^{2\beta\ei \sin^2\left(\phi\right)}\,\diff \phi\\
    &=2w\pi\sqrt{\frac{\ei}{\abs{\lambda}}}\e^{2\beta\ei}I^{\e}_0\left(\beta\ei\right)\,,
\end{align}
with $w$ being the number of periods until the initial spin configuration is reached again and $I^{\e}_v(z) = \e^{-\abs{\real(z)}}I_v(z)$ is the exponentially scaled modified Bessel function of first kind. Similarly, the contribution of one unstable quartic mode evaluates to
\begin{align}
    &\int_{-\infty}^{\infty}e^{\frac{\beta\abs{\lambda}}{2}p^2 - \frac{5\beta\abs{\lambda}^2}{144\ei}p^4}\diff p\\
    &= \frac{3\pi}{\sqrt{5}}\sqrt{\frac{\ei}{\abs{\lambda}}}e^{\frac{9}{5}\beta\ei}\left(I^e_{-\frac{1}{4}}\left(\frac{9}{10}\beta\ei\right)+I^e_{\frac{1}{4}}\left(\frac{9}{10}\beta\ei\right)\right)\,.
\end{align}
For a comparison of the dispersions used in the extended harmonic approximation with the ones explicitly computed using numerical mode following, see Supplementary Figure~S8.

\section*{}
\noindent{\large{\textbf{Acknowledgments}}\par
\normalsize
We thank Felix Nickel, Tim Drevelow, Soumyajyoti Haldar, André Kubetzka and Kirsten von Bergmann for valuable discussions. This research was supported in part through high-performance computing resources available at the Kiel University Computing Centre. We gratefully acknowledge financial support from the Deutsche Forschungsgemeinschaft (DFG, German Research Foundation) via project no.~418425860.
H.S. acknowledges financial support from the Icelandic Research Fund (grant No. 239435).

\section*{}
\noindent{\large{\textbf{Author contributions}}\par
\normalsize
L.K. performed the Monte Carlo simulations. L.K. and S.H. analyzed the data. L.K. and M.A.G. developed the extension for the harmonic approximation. B.B. and H.S. implemented important parts of the code regarding the higher-order interactions and the numerical mode following algorithms. L.K. and S.H. wrote the manuscript with contributions from all authors.

\section*{}
\noindent{\large{\textbf{Data availability}}\par
\normalsize
All relevant data are available from the corresponding authors upon reasonable request.

\section*{}
\noindent{\large{\textbf{Code availability}}\par
\normalsize
All relevant code is available from the corresponding authors upon reasonable request.

\section*{}
\noindent{\large{\textbf{Competing interests}}\par
\normalsize
The authors declare no competing interests.
\bibliography{references}

\begin{thebibliography}{47}%
\makeatletter
\providecommand \@ifxundefined [1]{%
 \@ifx{#1\undefined}
}%
\providecommand \@ifnum [1]{%
 \ifnum #1\expandafter \@firstoftwo
 \else \expandafter \@secondoftwo
 \fi
}%
\providecommand \@ifx [1]{%
 \ifx #1\expandafter \@firstoftwo
 \else \expandafter \@secondoftwo
 \fi
}%
\providecommand \natexlab [1]{#1}%
\providecommand \enquote  [1]{``#1''}%
\providecommand \bibnamefont  [1]{#1}%
\providecommand \bibfnamefont [1]{#1}%
\providecommand \citenamefont [1]{#1}%
\providecommand \href@noop [0]{\@secondoftwo}%
\providecommand \href [0]{\begingroup \@sanitize@url \@href}%
\providecommand \@href[1]{\@@startlink{#1}\@@href}%
\providecommand \@@href[1]{\endgroup#1\@@endlink}%
\providecommand \@sanitize@url [0]{\catcode `\\12\catcode `\$12\catcode `\&12\catcode `\#12\catcode `\^12\catcode `\_12\catcode `\%12\relax}%
\providecommand \@@startlink[1]{}%
\providecommand \@@endlink[0]{}%
\providecommand \url  [0]{\begingroup\@sanitize@url \@url }%
\providecommand \@url [1]{\endgroup\@href {#1}{\urlprefix }}%
\providecommand \urlprefix  [0]{URL }%
\providecommand \Eprint [0]{\href }%
\providecommand \doibase [0]{https://doi.org/}%
\providecommand \selectlanguage [0]{\@gobble}%
\providecommand \bibinfo  [0]{\@secondoftwo}%
\providecommand \bibfield  [0]{\@secondoftwo}%
\providecommand \translation [1]{[#1]}%
\providecommand \BibitemOpen [0]{}%
\providecommand \bibitemStop [0]{}%
\providecommand \bibitemNoStop [0]{.\EOS\space}%
\providecommand \EOS [0]{\spacefactor3000\relax}%
\providecommand \BibitemShut  [1]{\csname bibitem#1\endcsname}%
\let\auto@bib@innerbib\@empty
\bibitem [{\citenamefont {Baltz}\ \emph {et~al.}(2018)\citenamefont {Baltz}, \citenamefont {Manchon}, \citenamefont {Tsoi}, \citenamefont {Moriyama}, \citenamefont {Ono},\ and\ \citenamefont {Tserkovnyak}}]{baltz2018antiferromagnetic}%
  \BibitemOpen
  \bibfield  {author} {\bibinfo {author} {\bibfnamefont {V.}~\bibnamefont {Baltz}}, \bibinfo {author} {\bibfnamefont {A.}~\bibnamefont {Manchon}}, \bibinfo {author} {\bibfnamefont {M.}~\bibnamefont {Tsoi}}, \bibinfo {author} {\bibfnamefont {T.}~\bibnamefont {Moriyama}}, \bibinfo {author} {\bibfnamefont {T.}~\bibnamefont {Ono}},\ and\ \bibinfo {author} {\bibfnamefont {Y.}~\bibnamefont {Tserkovnyak}},\ }\bibfield  {title} {\bibinfo {title} {Antiferromagnetic spintronics},\ }\href@noop {} {\bibfield  {journal} {\bibinfo  {journal} {Rev. Mod. Phys.}\ }\textbf {\bibinfo {volume} {90}},\ \bibinfo {pages} {015005} (\bibinfo {year} {2018})}\BibitemShut {NoStop}%
\bibitem [{\citenamefont {Rimmler}\ \emph {et~al.}(2024)\citenamefont {Rimmler}, \citenamefont {Pal},\ and\ \citenamefont {Parkin}}]{Rimmler2024}%
  \BibitemOpen
  \bibfield  {author} {\bibinfo {author} {\bibfnamefont {B.}~\bibnamefont {Rimmler}}, \bibinfo {author} {\bibfnamefont {B.}~\bibnamefont {Pal}},\ and\ \bibinfo {author} {\bibfnamefont {S.}~\bibnamefont {Parkin}},\ }\bibfield  {title} {\bibinfo {title} {Non-collinear antiferromagnetic spintronics},\ }\href {https://doi.org/10.1038/s41578-024-00706-w} {\bibfield  {journal} {\bibinfo  {journal} {Nat. Rev. Mater.}\ } (\bibinfo {year} {2024})}\BibitemShut {NoStop}%
\bibitem [{\citenamefont {Qin}\ \emph {et~al.}(2023)\citenamefont {Qin}, \citenamefont {Yan}, \citenamefont {Wang}, \citenamefont {Chen}, \citenamefont {Meng}, \citenamefont {Dong}, \citenamefont {Zhu}, \citenamefont {Cai}, \citenamefont {Feng}, \citenamefont {Zhou}, \citenamefont {Liu}, \citenamefont {Zhang}, \citenamefont {Zeng}, \citenamefont {Zhang}, \citenamefont {Jiang},\ and\ \citenamefont {Liu}}]{Qin2023}%
  \BibitemOpen
  \bibfield  {author} {\bibinfo {author} {\bibfnamefont {P.}~\bibnamefont {Qin}}, \bibinfo {author} {\bibfnamefont {H.}~\bibnamefont {Yan}}, \bibinfo {author} {\bibfnamefont {X.}~\bibnamefont {Wang}}, \bibinfo {author} {\bibfnamefont {H.}~\bibnamefont {Chen}}, \bibinfo {author} {\bibfnamefont {Z.}~\bibnamefont {Meng}}, \bibinfo {author} {\bibfnamefont {J.}~\bibnamefont {Dong}}, \bibinfo {author} {\bibfnamefont {M.}~\bibnamefont {Zhu}}, \bibinfo {author} {\bibfnamefont {J.}~\bibnamefont {Cai}}, \bibinfo {author} {\bibfnamefont {Z.}~\bibnamefont {Feng}}, \bibinfo {author} {\bibfnamefont {X.}~\bibnamefont {Zhou}}, \bibinfo {author} {\bibfnamefont {L.}~\bibnamefont {Liu}}, \bibinfo {author} {\bibfnamefont {T.}~\bibnamefont {Zhang}}, \bibinfo {author} {\bibfnamefont {Z.}~\bibnamefont {Zeng}}, \bibinfo {author} {\bibfnamefont {J.}~\bibnamefont {Zhang}}, \bibinfo {author} {\bibfnamefont {C.}~\bibnamefont {Jiang}},\ and\ \bibinfo {author} {\bibfnamefont {Z.}~\bibnamefont {Liu}},\ }\bibfield  {title} {\bibinfo {title}
  {Room-temperature magnetoresistance in an all-antiferromagnetic tunnel junction},\ }\href@noop {} {\bibfield  {journal} {\bibinfo  {journal} {Nature}\ }\textbf {\bibinfo {volume} {485}},\ \bibinfo {pages} {613} (\bibinfo {year} {2023})}\BibitemShut {NoStop}%
\bibitem [{\citenamefont {Chen}\ \emph {et~al.}(2023)\citenamefont {Chen}, \citenamefont {Higo}, \citenamefont {Tanaka}, \citenamefont {Nomoto}, \citenamefont {Tsai}, \citenamefont {Idzuchi}, \citenamefont {Shiga}, \citenamefont {Sakamoto}, \citenamefont {Ando}, \citenamefont {Kosaki}, \citenamefont {Matsuo}, \citenamefont {Nishio-Hamane}, \citenamefont {Arita}, \citenamefont {Miwa},\ and\ \citenamefont {Nakatsuji}}]{Chen2023}%
  \BibitemOpen
  \bibfield  {author} {\bibinfo {author} {\bibfnamefont {X.}~\bibnamefont {Chen}}, \bibinfo {author} {\bibfnamefont {T.}~\bibnamefont {Higo}}, \bibinfo {author} {\bibfnamefont {K.}~\bibnamefont {Tanaka}}, \bibinfo {author} {\bibfnamefont {T.}~\bibnamefont {Nomoto}}, \bibinfo {author} {\bibfnamefont {H.}~\bibnamefont {Tsai}}, \bibinfo {author} {\bibfnamefont {H.}~\bibnamefont {Idzuchi}}, \bibinfo {author} {\bibfnamefont {M.}~\bibnamefont {Shiga}}, \bibinfo {author} {\bibfnamefont {S.}~\bibnamefont {Sakamoto}}, \bibinfo {author} {\bibfnamefont {R.}~\bibnamefont {Ando}}, \bibinfo {author} {\bibfnamefont {H.}~\bibnamefont {Kosaki}}, \bibinfo {author} {\bibfnamefont {T.}~\bibnamefont {Matsuo}}, \bibinfo {author} {\bibfnamefont {D.}~\bibnamefont {Nishio-Hamane}}, \bibinfo {author} {\bibfnamefont {R.}~\bibnamefont {Arita}}, \bibinfo {author} {\bibfnamefont {S.}~\bibnamefont {Miwa}},\ and\ \bibinfo {author} {\bibfnamefont {S.}~\bibnamefont {Nakatsuji}},\ }\bibfield  {title} {\bibinfo {title} {Octupole-driven
  magnetoresistance in an antiferromagnetic tunnel junction},\ }\href@noop {} {\bibfield  {journal} {\bibinfo  {journal} {Nature}\ }\textbf {\bibinfo {volume} {490}},\ \bibinfo {pages} {613} (\bibinfo {year} {2023})}\BibitemShut {NoStop}%
\bibitem [{\citenamefont {Zheng}\ \emph {et~al.}(2025)\citenamefont {Zheng}, \citenamefont {Jia}, \citenamefont {Zhang}, \citenamefont {Shen}, \citenamefont {Zhou}, \citenamefont {Cui}, \citenamefont {Ren}, \citenamefont {Chen}, \citenamefont {Jamaludin}, \citenamefont {Zhao}, \citenamefont {Xiao}, \citenamefont {Zhang}, \citenamefont {Du}, \citenamefont {Liu}, \citenamefont {Gradečak}, \citenamefont {Novoselov}, \citenamefont {Zhao}, \citenamefont {Xu}, \citenamefont {Zhang},\ and\ \citenamefont {Chen}}]{Zheng2025}%
  \BibitemOpen
  \bibfield  {author} {\bibinfo {author} {\bibfnamefont {Z.}~\bibnamefont {Zheng}}, \bibinfo {author} {\bibfnamefont {L.}~\bibnamefont {Jia}}, \bibinfo {author} {\bibfnamefont {Z.}~\bibnamefont {Zhang}}, \bibinfo {author} {\bibfnamefont {Q.}~\bibnamefont {Shen}}, \bibinfo {author} {\bibfnamefont {G.}~\bibnamefont {Zhou}}, \bibinfo {author} {\bibfnamefont {Z.}~\bibnamefont {Cui}}, \bibinfo {author} {\bibfnamefont {L.}~\bibnamefont {Ren}}, \bibinfo {author} {\bibfnamefont {Z.}~\bibnamefont {Chen}}, \bibinfo {author} {\bibfnamefont {N.~F.}\ \bibnamefont {Jamaludin}}, \bibinfo {author} {\bibfnamefont {T.}~\bibnamefont {Zhao}}, \bibinfo {author} {\bibfnamefont {R.}~\bibnamefont {Xiao}}, \bibinfo {author} {\bibfnamefont {Q.}~\bibnamefont {Zhang}}, \bibinfo {author} {\bibfnamefont {Y.}~\bibnamefont {Du}}, \bibinfo {author} {\bibfnamefont {L.}~\bibnamefont {Liu}}, \bibinfo {author} {\bibfnamefont {S.}~\bibnamefont {Gradečak}}, \bibinfo {author} {\bibfnamefont {K.~S.}\ \bibnamefont {Novoselov}}, \bibinfo {author}
  {\bibfnamefont {W.}~\bibnamefont {Zhao}}, \bibinfo {author} {\bibfnamefont {X.}~\bibnamefont {Xu}}, \bibinfo {author} {\bibfnamefont {Y.}~\bibnamefont {Zhang}},\ and\ \bibinfo {author} {\bibfnamefont {J.}~\bibnamefont {Chen}},\ }\bibfield  {title} {\bibinfo {title} {All-electrical perpendicular switching of chiral antiferromagnetic order},\ }\href@noop {} {\bibfield  {journal} {\bibinfo  {journal} {Nat. Mat.}\ } (\bibinfo {year} {2025})}\BibitemShut {NoStop}%
\bibitem [{\citenamefont {Kurz}\ \emph {et~al.}(2001)\citenamefont {Kurz}, \citenamefont {Bihlmayer}, \citenamefont {Hirai},\ and\ \citenamefont {Bl{\"u}gel}}]{kurz2001three}%
  \BibitemOpen
  \bibfield  {author} {\bibinfo {author} {\bibfnamefont {P.}~\bibnamefont {Kurz}}, \bibinfo {author} {\bibfnamefont {G.}~\bibnamefont {Bihlmayer}}, \bibinfo {author} {\bibfnamefont {K.}~\bibnamefont {Hirai}},\ and\ \bibinfo {author} {\bibfnamefont {S.}~\bibnamefont {Bl{\"u}gel}},\ }\bibfield  {title} {\bibinfo {title} {Three-dimensional spin structure on a two-dimensional lattice: {Mn/Cu(111)}},\ }\href@noop {} {\bibfield  {journal} {\bibinfo  {journal} {Phys. Rev. Lett.}\ }\textbf {\bibinfo {volume} {86}},\ \bibinfo {pages} {1106} (\bibinfo {year} {2001})}\BibitemShut {NoStop}%
\bibitem [{\citenamefont {Hanke}\ \emph {et~al.}(2016)\citenamefont {Hanke}, \citenamefont {Freimuth}, \citenamefont {Nandy}, \citenamefont {Zhang}, \citenamefont {Bl{\"u}gel},\ and\ \citenamefont {Mokrousov}}]{hanke2016role}%
  \BibitemOpen
  \bibfield  {author} {\bibinfo {author} {\bibfnamefont {J.-P.}\ \bibnamefont {Hanke}}, \bibinfo {author} {\bibfnamefont {F.}~\bibnamefont {Freimuth}}, \bibinfo {author} {\bibfnamefont {A.~K.}\ \bibnamefont {Nandy}}, \bibinfo {author} {\bibfnamefont {H.}~\bibnamefont {Zhang}}, \bibinfo {author} {\bibfnamefont {S.}~\bibnamefont {Bl{\"u}gel}},\ and\ \bibinfo {author} {\bibfnamefont {Y.}~\bibnamefont {Mokrousov}},\ }\bibfield  {title} {\bibinfo {title} {Role of {B}erry phase theory for describing orbital magnetism: From magnetic heterostructures to topological orbital ferromagnets},\ }\href@noop {} {\bibfield  {journal} {\bibinfo  {journal} {Phys. Rev. B}\ }\textbf {\bibinfo {volume} {94}},\ \bibinfo {pages} {121114} (\bibinfo {year} {2016})}\BibitemShut {NoStop}%
\bibitem [{\citenamefont {Haldar}\ \emph {et~al.}(2021)\citenamefont {Haldar}, \citenamefont {Meyer}, \citenamefont {Kubetzka},\ and\ \citenamefont {Heinze}}]{haldar2021distorted}%
  \BibitemOpen
  \bibfield  {author} {\bibinfo {author} {\bibfnamefont {S.}~\bibnamefont {Haldar}}, \bibinfo {author} {\bibfnamefont {S.}~\bibnamefont {Meyer}}, \bibinfo {author} {\bibfnamefont {A.}~\bibnamefont {Kubetzka}},\ and\ \bibinfo {author} {\bibfnamefont {S.}~\bibnamefont {Heinze}},\ }\bibfield  {title} {\bibinfo {title} {Distorted 3q state driven by topological-chiral magnetic interactions},\ }\href@noop {} {\bibfield  {journal} {\bibinfo  {journal} {Phys. Rev. B}\ }\textbf {\bibinfo {volume} {104}},\ \bibinfo {pages} {L180404} (\bibinfo {year} {2021})}\BibitemShut {NoStop}%
\bibitem [{\citenamefont {Martin}\ and\ \citenamefont {Batista}(2008)}]{martin2008itinerant}%
  \BibitemOpen
  \bibfield  {author} {\bibinfo {author} {\bibfnamefont {I.}~\bibnamefont {Martin}}\ and\ \bibinfo {author} {\bibfnamefont {C.}~\bibnamefont {Batista}},\ }\bibfield  {title} {\bibinfo {title} {Itinerant electron-driven chiral magnetic ordering and spontaneous quantum {H}all effect in triangular lattice models},\ }\href@noop {} {\bibfield  {journal} {\bibinfo  {journal} {Phys. Rev. Lett.}\ }\textbf {\bibinfo {volume} {101}},\ \bibinfo {pages} {156402} (\bibinfo {year} {2008})}\BibitemShut {NoStop}%
\bibitem [{\citenamefont {Takagi}\ \emph {et~al.}(2023)\citenamefont {Takagi}, \citenamefont {Takagi}, \citenamefont {Minami}, \citenamefont {Nomoto}, \citenamefont {Ohishi}, \citenamefont {Suzuki}, \citenamefont {Yanagi}, \citenamefont {Hirayama}, \citenamefont {Khanh}, \citenamefont {Karube} \emph {et~al.}}]{takagi2023spontaneous}%
  \BibitemOpen
  \bibfield  {author} {\bibinfo {author} {\bibfnamefont {H.}~\bibnamefont {Takagi}}, \bibinfo {author} {\bibfnamefont {R.}~\bibnamefont {Takagi}}, \bibinfo {author} {\bibfnamefont {S.}~\bibnamefont {Minami}}, \bibinfo {author} {\bibfnamefont {T.}~\bibnamefont {Nomoto}}, \bibinfo {author} {\bibfnamefont {K.}~\bibnamefont {Ohishi}}, \bibinfo {author} {\bibfnamefont {M.-T.}\ \bibnamefont {Suzuki}}, \bibinfo {author} {\bibfnamefont {Y.}~\bibnamefont {Yanagi}}, \bibinfo {author} {\bibfnamefont {M.}~\bibnamefont {Hirayama}}, \bibinfo {author} {\bibfnamefont {N.}~\bibnamefont {Khanh}}, \bibinfo {author} {\bibfnamefont {K.}~\bibnamefont {Karube}}, \emph {et~al.},\ }\bibfield  {title} {\bibinfo {title} {Spontaneous topological {H}all effect induced by non-coplanar antiferromagnetic order in intercalated van der {W}aals materials},\ }\href@noop {} {\bibfield  {journal} {\bibinfo  {journal} {Nat. Phys.}\ }\textbf {\bibinfo {volume} {19}},\ \bibinfo {pages} {961} (\bibinfo {year} {2023})}\BibitemShut {NoStop}%
\bibitem [{\citenamefont {Park}\ \emph {et~al.}(2023)\citenamefont {Park}, \citenamefont {Cho}, \citenamefont {Kim}, \citenamefont {An}, \citenamefont {Kang}, \citenamefont {Avdeev}, \citenamefont {Sibille}, \citenamefont {Iida}, \citenamefont {Kajimoto}, \citenamefont {Lee} \emph {et~al.}}]{park2023tetrahedral}%
  \BibitemOpen
  \bibfield  {author} {\bibinfo {author} {\bibfnamefont {P.}~\bibnamefont {Park}}, \bibinfo {author} {\bibfnamefont {W.}~\bibnamefont {Cho}}, \bibinfo {author} {\bibfnamefont {C.}~\bibnamefont {Kim}}, \bibinfo {author} {\bibfnamefont {Y.}~\bibnamefont {An}}, \bibinfo {author} {\bibfnamefont {Y.-G.}\ \bibnamefont {Kang}}, \bibinfo {author} {\bibfnamefont {M.}~\bibnamefont {Avdeev}}, \bibinfo {author} {\bibfnamefont {R.}~\bibnamefont {Sibille}}, \bibinfo {author} {\bibfnamefont {K.}~\bibnamefont {Iida}}, \bibinfo {author} {\bibfnamefont {R.}~\bibnamefont {Kajimoto}}, \bibinfo {author} {\bibfnamefont {K.~H.}\ \bibnamefont {Lee}}, \emph {et~al.},\ }\bibfield  {title} {\bibinfo {title} {Tetrahedral triple-q magnetic ordering and large spontaneous {H}all conductivity in the metallic triangular antiferromagnet {C}o$_{1/3}${TaS}$_2$},\ }\href@noop {} {\bibfield  {journal} {\bibinfo  {journal} {Nat. Commun.}\ }\textbf {\bibinfo {volume} {14}},\ \bibinfo {pages} {8346} (\bibinfo {year} {2023})}\BibitemShut {NoStop}%
\bibitem [{\citenamefont {Bedow}\ \emph {et~al.}(2020)\citenamefont {Bedow}, \citenamefont {Mascot}, \citenamefont {Posske}, \citenamefont {Uhrig}, \citenamefont {Wiesendanger}, \citenamefont {Rachel},\ and\ \citenamefont {Morr}}]{Bedow2020}%
  \BibitemOpen
  \bibfield  {author} {\bibinfo {author} {\bibfnamefont {J.}~\bibnamefont {Bedow}}, \bibinfo {author} {\bibfnamefont {E.}~\bibnamefont {Mascot}}, \bibinfo {author} {\bibfnamefont {T.}~\bibnamefont {Posske}}, \bibinfo {author} {\bibfnamefont {G.~S.}\ \bibnamefont {Uhrig}}, \bibinfo {author} {\bibfnamefont {R.}~\bibnamefont {Wiesendanger}}, \bibinfo {author} {\bibfnamefont {S.}~\bibnamefont {Rachel}},\ and\ \bibinfo {author} {\bibfnamefont {D.~K.}\ \bibnamefont {Morr}},\ }\bibfield  {title} {\bibinfo {title} {Topological superconductivity induced by a triple-q magnetic structure},\ }\href {https://doi.org/10.1103/PhysRevB.102.180504} {\bibfield  {journal} {\bibinfo  {journal} {Phys. Rev. B}\ }\textbf {\bibinfo {volume} {102}},\ \bibinfo {pages} {180504} (\bibinfo {year} {2020})}\BibitemShut {NoStop}%
\bibitem [{\citenamefont {Nickel}\ and\ \citenamefont {Heinze}(2025)}]{Nickel2025}%
  \BibitemOpen
  \bibfield  {author} {\bibinfo {author} {\bibfnamefont {F.}~\bibnamefont {Nickel}}\ and\ \bibinfo {author} {\bibfnamefont {S.}~\bibnamefont {Heinze}},\ }\bibfield  {title} {\bibinfo {title} {Topological properties of magnet-superconductor hybrid systems due to atomic-scale non-coplanar spin textures},\ }\href@noop {} {\bibfield  {journal} {\bibinfo  {journal} {npj Spintronics}\ }\textbf {\bibinfo {volume} {3}},\ \bibinfo {pages} {13} (\bibinfo {year} {2025})}\BibitemShut {NoStop}%
\bibitem [{\citenamefont {Spethmann}\ \emph {et~al.}(2020)\citenamefont {Spethmann}, \citenamefont {Meyer}, \citenamefont {von Bergmann}, \citenamefont {Wiesendanger}, \citenamefont {Heinze},\ and\ \citenamefont {Kubetzka}}]{spethmann2020discovery}%
  \BibitemOpen
  \bibfield  {author} {\bibinfo {author} {\bibfnamefont {J.}~\bibnamefont {Spethmann}}, \bibinfo {author} {\bibfnamefont {S.}~\bibnamefont {Meyer}}, \bibinfo {author} {\bibfnamefont {K.}~\bibnamefont {von Bergmann}}, \bibinfo {author} {\bibfnamefont {R.}~\bibnamefont {Wiesendanger}}, \bibinfo {author} {\bibfnamefont {S.}~\bibnamefont {Heinze}},\ and\ \bibinfo {author} {\bibfnamefont {A.}~\bibnamefont {Kubetzka}},\ }\bibfield  {title} {\bibinfo {title} {Discovery of magnetic single-and triple-q states in {Mn/Re}(0001)},\ }\href@noop {} {\bibfield  {journal} {\bibinfo  {journal} {Phys. Rev. Lett.}\ }\textbf {\bibinfo {volume} {124}},\ \bibinfo {pages} {227203} (\bibinfo {year} {2020})}\BibitemShut {NoStop}%
\bibitem [{\citenamefont {Nickel}\ \emph {et~al.}(2023)\citenamefont {Nickel}, \citenamefont {Kubetzka}, \citenamefont {Haldar}, \citenamefont {Wiesendanger}, \citenamefont {Heinze},\ and\ \citenamefont {von Bergmann}}]{nickel2023coupling}%
  \BibitemOpen
  \bibfield  {author} {\bibinfo {author} {\bibfnamefont {F.}~\bibnamefont {Nickel}}, \bibinfo {author} {\bibfnamefont {A.}~\bibnamefont {Kubetzka}}, \bibinfo {author} {\bibfnamefont {S.}~\bibnamefont {Haldar}}, \bibinfo {author} {\bibfnamefont {R.}~\bibnamefont {Wiesendanger}}, \bibinfo {author} {\bibfnamefont {S.}~\bibnamefont {Heinze}},\ and\ \bibinfo {author} {\bibfnamefont {K.}~\bibnamefont {von Bergmann}},\ }\bibfield  {title} {\bibinfo {title} {Coupling of the triple-q state to the atomic lattice by anisotropic symmetric exchange},\ }\href@noop {} {\bibfield  {journal} {\bibinfo  {journal} {Phys. Rev. B}\ }\textbf {\bibinfo {volume} {108}},\ \bibinfo {pages} {L180411} (\bibinfo {year} {2023})}\BibitemShut {NoStop}%
\bibitem [{\citenamefont {Saxena}\ \emph {et~al.}(2024)\citenamefont {Saxena}, \citenamefont {Gutzeit}, \citenamefont {Rodríguez-Sota}, \citenamefont {Haldar}, \citenamefont {Zahner}, \citenamefont {Wiesendanger}, \citenamefont {Kubetzka}, \citenamefont {Heinze},\ and\ \citenamefont {von Bergmann}}]{saxena2024}%
  \BibitemOpen
  \bibfield  {author} {\bibinfo {author} {\bibfnamefont {V.}~\bibnamefont {Saxena}}, \bibinfo {author} {\bibfnamefont {M.}~\bibnamefont {Gutzeit}}, \bibinfo {author} {\bibfnamefont {A.}~\bibnamefont {Rodríguez-Sota}}, \bibinfo {author} {\bibfnamefont {S.}~\bibnamefont {Haldar}}, \bibinfo {author} {\bibfnamefont {F.}~\bibnamefont {Zahner}}, \bibinfo {author} {\bibfnamefont {R.}~\bibnamefont {Wiesendanger}}, \bibinfo {author} {\bibfnamefont {A.}~\bibnamefont {Kubetzka}}, \bibinfo {author} {\bibfnamefont {S.}~\bibnamefont {Heinze}},\ and\ \bibinfo {author} {\bibfnamefont {K.}~\bibnamefont {von Bergmann}},\ }\href {https://arxiv.org/abs/2408.12580} {\bibinfo {title} {Strain-driven domain wall network with chiral junctions in an antiferromagnet}} (\bibinfo {year} {2024}),\ \Eprint {https://arxiv.org/abs/2408.12580} {arXiv:2408.12580} \BibitemShut {NoStop}%
\bibitem [{\citenamefont {Hoffmann}\ and\ \citenamefont {Bl{\"u}gel}(2020)}]{hoffmann2020systematic}%
  \BibitemOpen
  \bibfield  {author} {\bibinfo {author} {\bibfnamefont {M.}~\bibnamefont {Hoffmann}}\ and\ \bibinfo {author} {\bibfnamefont {S.}~\bibnamefont {Bl{\"u}gel}},\ }\bibfield  {title} {\bibinfo {title} {Systematic derivation of realistic spin models for beyond-{H}eisenberg solids},\ }\href@noop {} {\bibfield  {journal} {\bibinfo  {journal} {Phys. Rev. B}\ }\textbf {\bibinfo {volume} {101}},\ \bibinfo {pages} {024418} (\bibinfo {year} {2020})}\BibitemShut {NoStop}%
\bibitem [{\citenamefont {Beyer}\ \emph {et~al.}(2025)\citenamefont {Beyer}, \citenamefont {Gutzeit}, \citenamefont {Drevelow}, \citenamefont {Schwermer}, \citenamefont {Haldar},\ and\ \citenamefont {Heinze}}]{Beyer2025}%
  \BibitemOpen
  \bibfield  {author} {\bibinfo {author} {\bibfnamefont {B.}~\bibnamefont {Beyer}}, \bibinfo {author} {\bibfnamefont {M.}~\bibnamefont {Gutzeit}}, \bibinfo {author} {\bibfnamefont {T.}~\bibnamefont {Drevelow}}, \bibinfo {author} {\bibfnamefont {I.}~\bibnamefont {Schwermer}}, \bibinfo {author} {\bibfnamefont {S.}~\bibnamefont {Haldar}},\ and\ \bibinfo {author} {\bibfnamefont {S.}~\bibnamefont {Heinze}},\ }\href {https://arxiv.org/abs/2408.12580} {\bibinfo {title} {Bilayer triple-q state driven by interlayer higher-order exchange interactions}} (\bibinfo {year} {2025}),\ \Eprint {https://arxiv.org/abs/2506.05091} {arXiv:2506.05091} \BibitemShut {NoStop}%
\bibitem [{\citenamefont {Heinze}\ \emph {et~al.}(2011)\citenamefont {Heinze}, \citenamefont {von Bergmann}, \citenamefont {Menzel}, \citenamefont {Brede}, \citenamefont {Kubetzka}, \citenamefont {Wiesendanger}, \citenamefont {Bihlmayer},\ and\ \citenamefont {Bl{\"u}gel}}]{heinze2011spontaneous}%
  \BibitemOpen
  \bibfield  {author} {\bibinfo {author} {\bibfnamefont {S.}~\bibnamefont {Heinze}}, \bibinfo {author} {\bibfnamefont {K.}~\bibnamefont {von Bergmann}}, \bibinfo {author} {\bibfnamefont {M.}~\bibnamefont {Menzel}}, \bibinfo {author} {\bibfnamefont {J.}~\bibnamefont {Brede}}, \bibinfo {author} {\bibfnamefont {A.}~\bibnamefont {Kubetzka}}, \bibinfo {author} {\bibfnamefont {R.}~\bibnamefont {Wiesendanger}}, \bibinfo {author} {\bibfnamefont {G.}~\bibnamefont {Bihlmayer}},\ and\ \bibinfo {author} {\bibfnamefont {S.}~\bibnamefont {Bl{\"u}gel}},\ }\bibfield  {title} {\bibinfo {title} {Spontaneous atomic-scale magnetic skyrmion lattice in two dimensions},\ }\href@noop {} {\bibfield  {journal} {\bibinfo  {journal} {Nat. Phys.}\ }\textbf {\bibinfo {volume} {7}},\ \bibinfo {pages} {713} (\bibinfo {year} {2011})}\BibitemShut {NoStop}%
\bibitem [{\citenamefont {Yoshida}\ \emph {et~al.}(2012)\citenamefont {Yoshida}, \citenamefont {Schr\"oder}, \citenamefont {Ferriani}, \citenamefont {Serrate}, \citenamefont {Kubetzka}, \citenamefont {von Bergmann}, \citenamefont {Heinze},\ and\ \citenamefont {Wiesendanger}}]{Yoshida2012}%
  \BibitemOpen
  \bibfield  {author} {\bibinfo {author} {\bibfnamefont {Y.}~\bibnamefont {Yoshida}}, \bibinfo {author} {\bibfnamefont {S.}~\bibnamefont {Schr\"oder}}, \bibinfo {author} {\bibfnamefont {P.}~\bibnamefont {Ferriani}}, \bibinfo {author} {\bibfnamefont {D.}~\bibnamefont {Serrate}}, \bibinfo {author} {\bibfnamefont {A.}~\bibnamefont {Kubetzka}}, \bibinfo {author} {\bibfnamefont {K.}~\bibnamefont {von Bergmann}}, \bibinfo {author} {\bibfnamefont {S.}~\bibnamefont {Heinze}},\ and\ \bibinfo {author} {\bibfnamefont {R.}~\bibnamefont {Wiesendanger}},\ }\bibfield  {title} {\bibinfo {title} {Conical spin-spiral state in an ultrathin film driven by higher-order spin interactions},\ }\href {https://doi.org/10.1103/PhysRevLett.108.087205} {\bibfield  {journal} {\bibinfo  {journal} {Phys. Rev. Lett.}\ }\textbf {\bibinfo {volume} {108}},\ \bibinfo {pages} {087205} (\bibinfo {year} {2012})}\BibitemShut {NoStop}%
\bibitem [{\citenamefont {Romming}\ \emph {et~al.}(2018)\citenamefont {Romming}, \citenamefont {Pralow}, \citenamefont {Kubetzka}, \citenamefont {Hoffmann}, \citenamefont {von Malottki}, \citenamefont {Meyer}, \citenamefont {Dup\'e}, \citenamefont {Wiesendanger}, \citenamefont {von Bergmann},\ and\ \citenamefont {Heinze}}]{Romming2018}%
  \BibitemOpen
  \bibfield  {author} {\bibinfo {author} {\bibfnamefont {N.}~\bibnamefont {Romming}}, \bibinfo {author} {\bibfnamefont {H.}~\bibnamefont {Pralow}}, \bibinfo {author} {\bibfnamefont {A.}~\bibnamefont {Kubetzka}}, \bibinfo {author} {\bibfnamefont {M.}~\bibnamefont {Hoffmann}}, \bibinfo {author} {\bibfnamefont {S.}~\bibnamefont {von Malottki}}, \bibinfo {author} {\bibfnamefont {S.}~\bibnamefont {Meyer}}, \bibinfo {author} {\bibfnamefont {B.}~\bibnamefont {Dup\'e}}, \bibinfo {author} {\bibfnamefont {R.}~\bibnamefont {Wiesendanger}}, \bibinfo {author} {\bibfnamefont {K.}~\bibnamefont {von Bergmann}},\ and\ \bibinfo {author} {\bibfnamefont {S.}~\bibnamefont {Heinze}},\ }\bibfield  {title} {\bibinfo {title} {Competition of {D}zyaloshinskii-{M}oriya and higher-order exchange interactions in $\mathrm{Rh}/\mathrm{Fe}$ atomic bilayers on {I}r(111)},\ }\href {https://doi.org/10.1103/PhysRevLett.120.207201} {\bibfield  {journal} {\bibinfo  {journal} {Phys. Rev. Lett.}\ }\textbf {\bibinfo {volume} {120}},\ \bibinfo {pages}
  {207201} (\bibinfo {year} {2018})}\BibitemShut {NoStop}%
\bibitem [{\citenamefont {Kr\"onlein}\ \emph {et~al.}(2018)\citenamefont {Kr\"onlein}, \citenamefont {Schmitt}, \citenamefont {Hoffmann}, \citenamefont {Kemmer}, \citenamefont {Seubert}, \citenamefont {Vogt}, \citenamefont {K\"uspert}, \citenamefont {B\"ohme}, \citenamefont {Alonazi}, \citenamefont {K\"ugel}, \citenamefont {Albrithen}, \citenamefont {Bode}, \citenamefont {Bihlmayer},\ and\ \citenamefont {Bl\"ugel}}]{Kroenlein2018}%
  \BibitemOpen
  \bibfield  {author} {\bibinfo {author} {\bibfnamefont {A.}~\bibnamefont {Kr\"onlein}}, \bibinfo {author} {\bibfnamefont {M.}~\bibnamefont {Schmitt}}, \bibinfo {author} {\bibfnamefont {M.}~\bibnamefont {Hoffmann}}, \bibinfo {author} {\bibfnamefont {J.}~\bibnamefont {Kemmer}}, \bibinfo {author} {\bibfnamefont {N.}~\bibnamefont {Seubert}}, \bibinfo {author} {\bibfnamefont {M.}~\bibnamefont {Vogt}}, \bibinfo {author} {\bibfnamefont {J.}~\bibnamefont {K\"uspert}}, \bibinfo {author} {\bibfnamefont {M.}~\bibnamefont {B\"ohme}}, \bibinfo {author} {\bibfnamefont {B.}~\bibnamefont {Alonazi}}, \bibinfo {author} {\bibfnamefont {J.}~\bibnamefont {K\"ugel}}, \bibinfo {author} {\bibfnamefont {H.~A.}\ \bibnamefont {Albrithen}}, \bibinfo {author} {\bibfnamefont {M.}~\bibnamefont {Bode}}, \bibinfo {author} {\bibfnamefont {G.}~\bibnamefont {Bihlmayer}},\ and\ \bibinfo {author} {\bibfnamefont {S.}~\bibnamefont {Bl\"ugel}},\ }\bibfield  {title} {\bibinfo {title} {Magnetic ground state stabilized by three-site interactions:
  $\mathrm{Fe}/\mathrm{Rh}(111)$},\ }\href {https://doi.org/10.1103/PhysRevLett.120.207202} {\bibfield  {journal} {\bibinfo  {journal} {Phys. Rev. Lett.}\ }\textbf {\bibinfo {volume} {120}},\ \bibinfo {pages} {207202} (\bibinfo {year} {2018})}\BibitemShut {NoStop}%
\bibitem [{\citenamefont {Gutzeit}\ \emph {et~al.}(2022)\citenamefont {Gutzeit}, \citenamefont {Kubetzka}, \citenamefont {Haldar}, \citenamefont {Pralow}, \citenamefont {Goerzen}, \citenamefont {Wiesendanger}, \citenamefont {Heinze},\ and\ \citenamefont {von Bergmann}}]{gutzeit2022nano}%
  \BibitemOpen
  \bibfield  {author} {\bibinfo {author} {\bibfnamefont {M.}~\bibnamefont {Gutzeit}}, \bibinfo {author} {\bibfnamefont {A.}~\bibnamefont {Kubetzka}}, \bibinfo {author} {\bibfnamefont {S.}~\bibnamefont {Haldar}}, \bibinfo {author} {\bibfnamefont {H.}~\bibnamefont {Pralow}}, \bibinfo {author} {\bibfnamefont {M.~A.}\ \bibnamefont {Goerzen}}, \bibinfo {author} {\bibfnamefont {R.}~\bibnamefont {Wiesendanger}}, \bibinfo {author} {\bibfnamefont {S.}~\bibnamefont {Heinze}},\ and\ \bibinfo {author} {\bibfnamefont {K.}~\bibnamefont {von Bergmann}},\ }\bibfield  {title} {\bibinfo {title} {Nano-scale collinear multi-q states driven by higher-order interactions},\ }\href@noop {} {\bibfield  {journal} {\bibinfo  {journal} {Nat. Commun.}\ }\textbf {\bibinfo {volume} {13}},\ \bibinfo {pages} {5764} (\bibinfo {year} {2022})}\BibitemShut {NoStop}%
\bibitem [{\citenamefont {Gutzeit}\ \emph {et~al.}(2023)\citenamefont {Gutzeit}, \citenamefont {Drevelow}, \citenamefont {Goerzen}, \citenamefont {Haldar},\ and\ \citenamefont {Heinze}}]{gutzeit2023spontaneous}%
  \BibitemOpen
  \bibfield  {author} {\bibinfo {author} {\bibfnamefont {M.}~\bibnamefont {Gutzeit}}, \bibinfo {author} {\bibfnamefont {T.}~\bibnamefont {Drevelow}}, \bibinfo {author} {\bibfnamefont {M.~A.}\ \bibnamefont {Goerzen}}, \bibinfo {author} {\bibfnamefont {S.}~\bibnamefont {Haldar}},\ and\ \bibinfo {author} {\bibfnamefont {S.}~\bibnamefont {Heinze}},\ }\bibfield  {title} {\bibinfo {title} {Spontaneous square versus hexagonal nanoscale skyrmion lattices in {Fe/Ir}(111)},\ }\href@noop {} {\bibfield  {journal} {\bibinfo  {journal} {Phys. Rev. B}\ }\textbf {\bibinfo {volume} {108}},\ \bibinfo {pages} {L060405} (\bibinfo {year} {2023})}\BibitemShut {NoStop}%
\bibitem [{\citenamefont {Nickel}\ \emph {et~al.}(2025)\citenamefont {Nickel}, \citenamefont {Kubetzka}, \citenamefont {Gutzeit}, \citenamefont {Wiesendanger}, \citenamefont {von Bergmann},\ and\ \citenamefont {Heinze}}]{Nickel2025b}%
  \BibitemOpen
  \bibfield  {author} {\bibinfo {author} {\bibfnamefont {F.}~\bibnamefont {Nickel}}, \bibinfo {author} {\bibfnamefont {A.}~\bibnamefont {Kubetzka}}, \bibinfo {author} {\bibfnamefont {M.}~\bibnamefont {Gutzeit}}, \bibinfo {author} {\bibfnamefont {R.}~\bibnamefont {Wiesendanger}}, \bibinfo {author} {\bibfnamefont {K.}~\bibnamefont {von Bergmann}},\ and\ \bibinfo {author} {\bibfnamefont {S.}~\bibnamefont {Heinze}},\ }\bibfield  {title} {\bibinfo {title} {Antiferromagnetic order of topological orbital moments in atomic-scale skyrmion lattices},\ }\href@noop {} {\bibfield  {journal} {\bibinfo  {journal} {npj Spintronics}\ }\textbf {\bibinfo {volume} {3}},\ \bibinfo {pages} {7} (\bibinfo {year} {2025})}\BibitemShut {NoStop}%
\bibitem [{\citenamefont {Xu}\ \emph {et~al.}(2022)\citenamefont {Xu}, \citenamefont {Li}, \citenamefont {Chen}, \citenamefont {Zhang}, \citenamefont {Xiang},\ and\ \citenamefont {Bellaiche}}]{Xu2022}%
  \BibitemOpen
  \bibfield  {author} {\bibinfo {author} {\bibfnamefont {C.}~\bibnamefont {Xu}}, \bibinfo {author} {\bibfnamefont {X.}~\bibnamefont {Li}}, \bibinfo {author} {\bibfnamefont {P.}~\bibnamefont {Chen}}, \bibinfo {author} {\bibfnamefont {Y.}~\bibnamefont {Zhang}}, \bibinfo {author} {\bibfnamefont {H.}~\bibnamefont {Xiang}},\ and\ \bibinfo {author} {\bibfnamefont {L.}~\bibnamefont {Bellaiche}},\ }\bibfield  {title} {\bibinfo {title} {Assembling diverse skyrmionic phases in {Fe$_3$GeTe$_2$} monolayers},\ }\href {https://doi.org/https://doi.org/10.1002/adma.202107779} {\bibfield  {journal} {\bibinfo  {journal} {Advanced Materials}\ }\textbf {\bibinfo {volume} {34}},\ \bibinfo {pages} {2107779} (\bibinfo {year} {2022})}\BibitemShut {NoStop}%
\bibitem [{\citenamefont {Li}\ \emph {et~al.}(2023)\citenamefont {Li}, \citenamefont {Yu}, \citenamefont {Liang}, \citenamefont {Ga},\ and\ \citenamefont {Yang}}]{Li2023}%
  \BibitemOpen
  \bibfield  {author} {\bibinfo {author} {\bibfnamefont {P.}~\bibnamefont {Li}}, \bibinfo {author} {\bibfnamefont {D.}~\bibnamefont {Yu}}, \bibinfo {author} {\bibfnamefont {J.}~\bibnamefont {Liang}}, \bibinfo {author} {\bibfnamefont {Y.}~\bibnamefont {Ga}},\ and\ \bibinfo {author} {\bibfnamefont {H.}~\bibnamefont {Yang}},\ }\bibfield  {title} {\bibinfo {title} {Topological spin textures in {$1T$}-phase {J}anus magnets: Interplay between {Dzyaloshinskii-Moriya} interaction, magnetic frustration, and isotropic higher-order interactions},\ }\href {https://doi.org/10.1103/PhysRevB.107.054408} {\bibfield  {journal} {\bibinfo  {journal} {Phys. Rev. B}\ }\textbf {\bibinfo {volume} {107}},\ \bibinfo {pages} {054408} (\bibinfo {year} {2023})}\BibitemShut {NoStop}%
\bibitem [{\citenamefont {Pan}\ \emph {et~al.}(2024)\citenamefont {Pan}, \citenamefont {Xu}, \citenamefont {Li}, \citenamefont {Xu}, \citenamefont {Liu}, \citenamefont {Gu},\ and\ \citenamefont {Duan}}]{Pan2024}%
  \BibitemOpen
  \bibfield  {author} {\bibinfo {author} {\bibfnamefont {W.}~\bibnamefont {Pan}}, \bibinfo {author} {\bibfnamefont {C.}~\bibnamefont {Xu}}, \bibinfo {author} {\bibfnamefont {X.}~\bibnamefont {Li}}, \bibinfo {author} {\bibfnamefont {Z.}~\bibnamefont {Xu}}, \bibinfo {author} {\bibfnamefont {B.}~\bibnamefont {Liu}}, \bibinfo {author} {\bibfnamefont {B.-L.}\ \bibnamefont {Gu}},\ and\ \bibinfo {author} {\bibfnamefont {W.}~\bibnamefont {Duan}},\ }\bibfield  {title} {\bibinfo {title} {Chiral magnetism in lithium-decorated monolayer {${\mathrm{CrTe}}_{2}$}: Interplay between {Dzyaloshinskii-Moriya} interaction and higher-order interactions},\ }\href {https://doi.org/10.1103/PhysRevB.109.214405} {\bibfield  {journal} {\bibinfo  {journal} {Phys. Rev. B}\ }\textbf {\bibinfo {volume} {109}},\ \bibinfo {pages} {214405} (\bibinfo {year} {2024})}\BibitemShut {NoStop}%
\bibitem [{\citenamefont {Wieser}\ \emph {et~al.}(2008)\citenamefont {Wieser}, \citenamefont {Vedmedenko},\ and\ \citenamefont {Wiesendanger}}]{wieser2008entropy}%
  \BibitemOpen
  \bibfield  {author} {\bibinfo {author} {\bibfnamefont {R.}~\bibnamefont {Wieser}}, \bibinfo {author} {\bibfnamefont {E.}~\bibnamefont {Vedmedenko}},\ and\ \bibinfo {author} {\bibfnamefont {R.}~\bibnamefont {Wiesendanger}},\ }\bibfield  {title} {\bibinfo {title} {Entropy driven phase transition in itinerant antiferromagnetic monolayers},\ }\href@noop {} {\bibfield  {journal} {\bibinfo  {journal} {Phys. Rev. B}\ }\textbf {\bibinfo {volume} {77}},\ \bibinfo {pages} {064410} (\bibinfo {year} {2008})}\BibitemShut {NoStop}%
\bibitem [{\citenamefont {Bessarab}\ \emph {et~al.}(2012)\citenamefont {Bessarab}, \citenamefont {Uzdin},\ and\ \citenamefont {J{\'o}nsson}}]{bessarab2012harmonic}%
  \BibitemOpen
  \bibfield  {author} {\bibinfo {author} {\bibfnamefont {P.~F.}\ \bibnamefont {Bessarab}}, \bibinfo {author} {\bibfnamefont {V.~M.}\ \bibnamefont {Uzdin}},\ and\ \bibinfo {author} {\bibfnamefont {H.}~\bibnamefont {J{\'o}nsson}},\ }\bibfield  {title} {\bibinfo {title} {Harmonic transition-state theory of thermal spin transitions},\ }\href@noop {} {\bibfield  {journal} {\bibinfo  {journal} {Phys. Rev. B}\ }\textbf {\bibinfo {volume} {85}},\ \bibinfo {pages} {184409} (\bibinfo {year} {2012})}\BibitemShut {NoStop}%
\bibitem [{\citenamefont {Goerzen}\ \emph {et~al.}(2022)\citenamefont {Goerzen}, \citenamefont {von Malottki}, \citenamefont {Kwiatkowski}, \citenamefont {Bessarab},\ and\ \citenamefont {Heinze}}]{goerzen2022atomistic}%
  \BibitemOpen
  \bibfield  {author} {\bibinfo {author} {\bibfnamefont {M.~A.}\ \bibnamefont {Goerzen}}, \bibinfo {author} {\bibfnamefont {S.}~\bibnamefont {von Malottki}}, \bibinfo {author} {\bibfnamefont {G.~J.}\ \bibnamefont {Kwiatkowski}}, \bibinfo {author} {\bibfnamefont {P.~F.}\ \bibnamefont {Bessarab}},\ and\ \bibinfo {author} {\bibfnamefont {S.}~\bibnamefont {Heinze}},\ }\bibfield  {title} {\bibinfo {title} {Atomistic spin simulations of electric-field-assisted nucleation and annihilation of magnetic skyrmions in {Pd/Fe/Ir}(111)},\ }\href@noop {} {\bibfield  {journal} {\bibinfo  {journal} {Phys. Rev. B}\ }\textbf {\bibinfo {volume} {105}},\ \bibinfo {pages} {214435} (\bibinfo {year} {2022})}\BibitemShut {NoStop}%
\bibitem [{\citenamefont {Lobanov}\ \emph {et~al.}(2016)\citenamefont {Lobanov}, \citenamefont {J\'onsson},\ and\ \citenamefont {Uzdin}}]{Lobanov2016}%
  \BibitemOpen
  \bibfield  {author} {\bibinfo {author} {\bibfnamefont {I.~S.}\ \bibnamefont {Lobanov}}, \bibinfo {author} {\bibfnamefont {H.}~\bibnamefont {J\'onsson}},\ and\ \bibinfo {author} {\bibfnamefont {V.~M.}\ \bibnamefont {Uzdin}},\ }\bibfield  {title} {\bibinfo {title} {Mechanism and activation energy of magnetic skyrmion annihilation obtained from minimum energy path calculations},\ }\href {https://doi.org/10.1103/PhysRevB.94.174418} {\bibfield  {journal} {\bibinfo  {journal} {Phys. Rev. B}\ }\textbf {\bibinfo {volume} {94}},\ \bibinfo {pages} {174418} (\bibinfo {year} {2016})}\BibitemShut {NoStop}%
\bibitem [{\citenamefont {Draaisma}\ and\ \citenamefont {de~Jonge}(1988)}]{Draaisma1988}%
  \BibitemOpen
  \bibfield  {author} {\bibinfo {author} {\bibfnamefont {H.~J.~G.}\ \bibnamefont {Draaisma}}\ and\ \bibinfo {author} {\bibfnamefont {W.~J.~M.}\ \bibnamefont {de~Jonge}},\ }\bibfield  {title} {\bibinfo {title} {Surface and volume anisotropy from dipole‐dipole interactions in ultrathin ferromagnetic films},\ }\href {https://doi.org/10.1063/1.341397} {\bibfield  {journal} {\bibinfo  {journal} {J. Appl. Phys.}\ }\textbf {\bibinfo {volume} {64}},\ \bibinfo {pages} {3610} (\bibinfo {year} {1988})}\BibitemShut {NoStop}%
\bibitem [{\citenamefont {Gvozdikova}\ and\ \citenamefont {Zhitomirsky}(2005)}]{gvozdikova2005monte}%
  \BibitemOpen
  \bibfield  {author} {\bibinfo {author} {\bibfnamefont {M.}~\bibnamefont {Gvozdikova}}\ and\ \bibinfo {author} {\bibfnamefont {M.}~\bibnamefont {Zhitomirsky}},\ }\bibfield  {title} {\bibinfo {title} {A {Monte Carlo} study of the first-order transition in a {H}eisenberg fcc antiferromagnet},\ }\href@noop {} {\bibfield  {journal} {\bibinfo  {journal} {JETP Letters}\ }\textbf {\bibinfo {volume} {81}},\ \bibinfo {pages} {236} (\bibinfo {year} {2005})}\BibitemShut {NoStop}%
\bibitem [{\citenamefont {R{\'o}zsa}\ \emph {et~al.}(2016)\citenamefont {R{\'o}zsa}, \citenamefont {Simon}, \citenamefont {Palot{\'a}s}, \citenamefont {Udvardi},\ and\ \citenamefont {Szunyogh}}]{rozsa2016complex}%
  \BibitemOpen
  \bibfield  {author} {\bibinfo {author} {\bibfnamefont {L.}~\bibnamefont {R{\'o}zsa}}, \bibinfo {author} {\bibfnamefont {E.}~\bibnamefont {Simon}}, \bibinfo {author} {\bibfnamefont {K.}~\bibnamefont {Palot{\'a}s}}, \bibinfo {author} {\bibfnamefont {L.}~\bibnamefont {Udvardi}},\ and\ \bibinfo {author} {\bibfnamefont {L.}~\bibnamefont {Szunyogh}},\ }\bibfield  {title} {\bibinfo {title} {Complex magnetic phase diagram and skyrmion lifetime in an ultrathin film from atomistic simulations},\ }\href@noop {} {\bibfield  {journal} {\bibinfo  {journal} {Phys. Rev. B}\ }\textbf {\bibinfo {volume} {93}},\ \bibinfo {pages} {024417} (\bibinfo {year} {2016})}\BibitemShut {NoStop}%
\bibitem [{\citenamefont {Nocedal}\ and\ \citenamefont {Wright}(1999)}]{nocedal1999numerical}%
  \BibitemOpen
  \bibfield  {author} {\bibinfo {author} {\bibfnamefont {J.}~\bibnamefont {Nocedal}}\ and\ \bibinfo {author} {\bibfnamefont {S.~J.}\ \bibnamefont {Wright}},\ }\href@noop {} {\emph {\bibinfo {title} {Numerical optimization}}}\ (\bibinfo  {publisher} {Springer},\ \bibinfo {year} {1999})\BibitemShut {NoStop}%
\bibitem [{\citenamefont {Ivanov}\ \emph {et~al.}(2021)\citenamefont {Ivanov}, \citenamefont {Uzdin},\ and\ \citenamefont {J{\'o}nsson}}]{ivanov2021fast}%
  \BibitemOpen
  \bibfield  {author} {\bibinfo {author} {\bibfnamefont {A.~V.}\ \bibnamefont {Ivanov}}, \bibinfo {author} {\bibfnamefont {V.~M.}\ \bibnamefont {Uzdin}},\ and\ \bibinfo {author} {\bibfnamefont {H.}~\bibnamefont {J{\'o}nsson}},\ }\bibfield  {title} {\bibinfo {title} {Fast and robust algorithm for energy minimization of spin systems applied in an analysis of high temperature spin configurations in terms of skyrmion density},\ }\href@noop {} {\bibfield  {journal} {\bibinfo  {journal} {Comput. Phys. Commun.}\ }\textbf {\bibinfo {volume} {260}},\ \bibinfo {pages} {107749} (\bibinfo {year} {2021})}\BibitemShut {NoStop}%
\bibitem [{\citenamefont {Goerzen}(2024)}]{goerzen2024PhD}%
  \BibitemOpen
  \bibfield  {author} {\bibinfo {author} {\bibfnamefont {M.}~\bibnamefont {Goerzen}},\ }\emph {\bibinfo {title} {Thermal equilibrium and stability of complementary topological solitons in two-dimensional magnets}},\ \href@noop {} {Ph.D. thesis},\ \bibinfo  {school} {Kiel University} (\bibinfo {year} {2024})\BibitemShut {NoStop}%
\bibitem [{\citenamefont {M{\"u}ller}\ \emph {et~al.}(2018)\citenamefont {M{\"u}ller}, \citenamefont {Bessarab}, \citenamefont {Vlasov}, \citenamefont {Lux}, \citenamefont {Kiselev}, \citenamefont {Bl{\"u}gel}, \citenamefont {Uzdin},\ and\ \citenamefont {J{\'o}nsson}}]{muller2018duplication}%
  \BibitemOpen
  \bibfield  {author} {\bibinfo {author} {\bibfnamefont {G.~P.}\ \bibnamefont {M{\"u}ller}}, \bibinfo {author} {\bibfnamefont {P.~F.}\ \bibnamefont {Bessarab}}, \bibinfo {author} {\bibfnamefont {S.~M.}\ \bibnamefont {Vlasov}}, \bibinfo {author} {\bibfnamefont {F.}~\bibnamefont {Lux}}, \bibinfo {author} {\bibfnamefont {N.~S.}\ \bibnamefont {Kiselev}}, \bibinfo {author} {\bibfnamefont {S.}~\bibnamefont {Bl{\"u}gel}}, \bibinfo {author} {\bibfnamefont {V.~M.}\ \bibnamefont {Uzdin}},\ and\ \bibinfo {author} {\bibfnamefont {H.}~\bibnamefont {J{\'o}nsson}},\ }\bibfield  {title} {\bibinfo {title} {Duplication, collapse, and escape of magnetic skyrmions revealed using a systematic saddle point search method},\ }\href@noop {} {\bibfield  {journal} {\bibinfo  {journal} {Phys. Rev. Lett.}\ }\textbf {\bibinfo {volume} {121}},\ \bibinfo {pages} {197202} (\bibinfo {year} {2018})}\BibitemShut {NoStop}%
\bibitem [{\citenamefont {von Malottki}\ \emph {et~al.}(2025)\citenamefont {von Malottki}, \citenamefont {Goerzen}, \citenamefont {Schrautzer}, \citenamefont {Bessarab},\ and\ \citenamefont {Heinze}}]{malottki2025eigenmode}%
  \BibitemOpen
  \bibfield  {author} {\bibinfo {author} {\bibfnamefont {S.}~\bibnamefont {von Malottki}}, \bibinfo {author} {\bibfnamefont {M.~A.}\ \bibnamefont {Goerzen}}, \bibinfo {author} {\bibfnamefont {H.}~\bibnamefont {Schrautzer}}, \bibinfo {author} {\bibfnamefont {P.~F.}\ \bibnamefont {Bessarab}},\ and\ \bibinfo {author} {\bibfnamefont {S.}~\bibnamefont {Heinze}},\ }\bibfield  {title} {\bibinfo {title} {Eigenmode following for direct entropy calculation and characterization of magnetic systems},\ }\href@noop {} {\bibfield  {journal} {\bibinfo  {journal} {arXiv preprint arXiv:2503.12109}\ } (\bibinfo {year} {2025})}\BibitemShut {NoStop}%
\bibitem [{\citenamefont {Henley}(1989)}]{henley1989ordering}%
  \BibitemOpen
  \bibfield  {author} {\bibinfo {author} {\bibfnamefont {C.~L.}\ \bibnamefont {Henley}},\ }\bibfield  {title} {\bibinfo {title} {Ordering due to disorder in a frustrated vector antiferromagnet},\ }\href@noop {} {\bibfield  {journal} {\bibinfo  {journal} {Phys. Rev. Lett.}\ }\textbf {\bibinfo {volume} {62}},\ \bibinfo {pages} {2056} (\bibinfo {year} {1989})}\BibitemShut {NoStop}%
\bibitem [{\citenamefont {Grytsiuk}\ \emph {et~al.}(2020)\citenamefont {Grytsiuk}, \citenamefont {Hanke}, \citenamefont {Hoffmann}, \citenamefont {Bouaziz}, \citenamefont {Gomonay}, \citenamefont {Bihlmayer}, \citenamefont {Lounis}, \citenamefont {Mokrousov},\ and\ \citenamefont {Bl{\"{u}}gel}}]{Grytsiuk2020}%
  \BibitemOpen
  \bibfield  {author} {\bibinfo {author} {\bibfnamefont {S.}~\bibnamefont {Grytsiuk}}, \bibinfo {author} {\bibfnamefont {J.-P.}\ \bibnamefont {Hanke}}, \bibinfo {author} {\bibfnamefont {M.}~\bibnamefont {Hoffmann}}, \bibinfo {author} {\bibfnamefont {J.}~\bibnamefont {Bouaziz}}, \bibinfo {author} {\bibfnamefont {O.}~\bibnamefont {Gomonay}}, \bibinfo {author} {\bibfnamefont {G.}~\bibnamefont {Bihlmayer}}, \bibinfo {author} {\bibfnamefont {S.}~\bibnamefont {Lounis}}, \bibinfo {author} {\bibfnamefont {Y.}~\bibnamefont {Mokrousov}},\ and\ \bibinfo {author} {\bibfnamefont {S.}~\bibnamefont {Bl{\"{u}}gel}},\ }\bibfield  {title} {\bibinfo {title} {{Topological–chiral magnetic interactions driven by emergent orbital magnetism}},\ }\href {https://doi.org/10.1038/s41467-019-14030-3} {\bibfield  {journal} {\bibinfo  {journal} {Nat. Commun.}\ }\textbf {\bibinfo {volume} {11}},\ \bibinfo {pages} {511} (\bibinfo {year} {2020})}\BibitemShut {NoStop}%
\bibitem [{\citenamefont {Kurz}(2000)}]{Kurz_Phd_thesis}%
  \BibitemOpen
  \bibfield  {author} {\bibinfo {author} {\bibfnamefont {P.}~\bibnamefont {Kurz}},\ }\emph {\bibinfo {title} {{N}on-collinear magnetism at surfaces and in ultrathin films}},\ \href@noop {} {Ph.D. thesis},\ \bibinfo  {school} {RWTH Aachen} (\bibinfo {year} {2000})\BibitemShut {NoStop}%
\bibitem [{\citenamefont {Spethmann}\ \emph {et~al.}(2021)\citenamefont {Spethmann}, \citenamefont {Gr{\"u}nebohm}, \citenamefont {Wiesendanger}, \citenamefont {von Bergmann},\ and\ \citenamefont {Kubetzka}}]{spethmann2021discovery}%
  \BibitemOpen
  \bibfield  {author} {\bibinfo {author} {\bibfnamefont {J.}~\bibnamefont {Spethmann}}, \bibinfo {author} {\bibfnamefont {M.}~\bibnamefont {Gr{\"u}nebohm}}, \bibinfo {author} {\bibfnamefont {R.}~\bibnamefont {Wiesendanger}}, \bibinfo {author} {\bibfnamefont {K.}~\bibnamefont {von Bergmann}},\ and\ \bibinfo {author} {\bibfnamefont {A.}~\bibnamefont {Kubetzka}},\ }\bibfield  {title} {\bibinfo {title} {Discovery and characterization of a new type of domain wall in a row-wise antiferromagnet},\ }\href@noop {} {\bibfield  {journal} {\bibinfo  {journal} {Nat. Commun.}\ }\textbf {\bibinfo {volume} {12}},\ \bibinfo {pages} {3488} (\bibinfo {year} {2021})}\BibitemShut {NoStop}%
\bibitem [{\citenamefont {Zahner}\ \emph {et~al.}(2025)\citenamefont {Zahner}, \citenamefont {Haldar}, \citenamefont {Wiesendanger}, \citenamefont {Heinze}, \citenamefont {von Bergmann},\ and\ \citenamefont {Kubetzka}}]{zahner2024kicking}%
  \BibitemOpen
  \bibfield  {author} {\bibinfo {author} {\bibfnamefont {F.}~\bibnamefont {Zahner}}, \bibinfo {author} {\bibfnamefont {S.}~\bibnamefont {Haldar}}, \bibinfo {author} {\bibfnamefont {R.}~\bibnamefont {Wiesendanger}}, \bibinfo {author} {\bibfnamefont {S.}~\bibnamefont {Heinze}}, \bibinfo {author} {\bibfnamefont {K.}~\bibnamefont {von Bergmann}},\ and\ \bibinfo {author} {\bibfnamefont {A.}~\bibnamefont {Kubetzka}},\ }\bibfield  {title} {\bibinfo {title} {Anisotropic atom motion on a row-wise antiferromagnetic surface},\ }\href@noop {} {\bibfield  {journal} {\bibinfo  {journal} {Nat. Commun.}\ }\textbf {\bibinfo {volume} {16}},\ \bibinfo {pages} {4942} (\bibinfo {year} {2025})}\BibitemShut {NoStop}%
\bibitem [{\citenamefont {Sessi}\ \emph {et~al.}(2009)\citenamefont {Sessi}, \citenamefont {Guisinger}, \citenamefont {Guest},\ and\ \citenamefont {Bode}}]{Sessi2009}%
  \BibitemOpen
  \bibfield  {author} {\bibinfo {author} {\bibfnamefont {P.}~\bibnamefont {Sessi}}, \bibinfo {author} {\bibfnamefont {N.~P.}\ \bibnamefont {Guisinger}}, \bibinfo {author} {\bibfnamefont {J.~R.}\ \bibnamefont {Guest}},\ and\ \bibinfo {author} {\bibfnamefont {M.}~\bibnamefont {Bode}},\ }\bibfield  {title} {\bibinfo {title} {Temperature and size dependence of antiferromagnetism in {Mn} nanostructures},\ }\href {https://doi.org/10.1103/PhysRevLett.103.167201} {\bibfield  {journal} {\bibinfo  {journal} {Phys. Rev. Lett.}\ }\textbf {\bibinfo {volume} {103}},\ \bibinfo {pages} {167201} (\bibinfo {year} {2009})}\BibitemShut {NoStop}%
\bibitem [{\citenamefont {Alzate-Cardona}\ \emph {et~al.}(2019)\citenamefont {Alzate-Cardona}, \citenamefont {Sabogal-Su{\'a}rez}, \citenamefont {Evans},\ and\ \citenamefont {Restrepo-Parra}}]{alzate2019optimal}%
  \BibitemOpen
  \bibfield  {author} {\bibinfo {author} {\bibfnamefont {J.}~\bibnamefont {Alzate-Cardona}}, \bibinfo {author} {\bibfnamefont {D.}~\bibnamefont {Sabogal-Su{\'a}rez}}, \bibinfo {author} {\bibfnamefont {R.}~\bibnamefont {Evans}},\ and\ \bibinfo {author} {\bibfnamefont {E.}~\bibnamefont {Restrepo-Parra}},\ }\bibfield  {title} {\bibinfo {title} {Optimal phase space sampling for {Monte Carlo} simulations of {H}eisenberg spin systems},\ }\href@noop {} {\bibfield  {journal} {\bibinfo  {journal} {J. Phys. Condens. Matter}\ }\textbf {\bibinfo {volume} {31}},\ \bibinfo {pages} {095802} (\bibinfo {year} {2019})}\BibitemShut {NoStop}%
\end{thebibliography}%
\end{document}